\documentclass[a4paper,11pt]{article}
\usepackage{jheppub}
\usepackage{graphicx}  
\usepackage{amsmath, amssymb}   
\usepackage{dsfont} 
\usepackage{amsfonts}
\usepackage{hyperref}
\usepackage[applemac]{inputenc}
\usepackage[british]{babel}

\newlength{\wth}
\newcommand{\be}{\begin{equation}}
\newcommand{\ee}{\end{equation}}
\newcommand{\bea}{\begin{eqnarray}}
\newcommand{\eea}{\end{eqnarray}}

\newcommand{\beqa}{\begin{eqnarray}}
\newcommand{\eeqa}{\end{eqnarray}}

\newcommand{\del}{\partial}

\newcommand{\lp}{\left(}
\newcommand{\rp}{\right)}
\newcommand{\ls}{\left[}
\newcommand{\rs}{\right]}

\renewcommand{\t}{\tau}

\newcommand{\V}{\mathcal{V}}
\newcommand{\Vmin}{{\mathcal V}_{\rm min}}
\renewcommand{\O}{\mathcal{O}}
\def\K{K{\"a}hler}
\def\ib{{\bar \imath}}
\def\jb{{\bar \jmath}}
\def\Ib{{\bar I}}
\def\Jb{{\bar J}}

\newcommand{\Pv}{\ensuremath{\vec{\Pi}}}
\newcommand{\G}{\ensuremath{{\cal G}}}

\newcommand{\Kcs}{\ensuremath{K_{\rm c.s.}}}
\newcommand{\Kk}{\ensuremath{K_{\rm K}}}
\newcommand{\Ks}{\ensuremath{K_{\rm dil}}}
\renewcommand{\Re}{{\rm Re}}
\renewcommand{\Im}{{\rm Im}}

\newcommand{\beq}{\begin{equation}}
\newcommand{\eeq}{\end{equation}}

\def\rmi{{\rm i}}

\newcommand{\bW}{\ensuremath{\overline W}}

\newcommand{\bF}{\ensuremath{\bar F}}

\newcommand{\bZ}{\ensuremath{\overline Z}}

\usepackage{tikz}
\usetikzlibrary{fit}
\tikzset{%
	highlight/.style={rectangle,rounded corners,fill=red!15,draw,
		fill opacity=0.5,thick,inner sep=0pt}
}

\numberwithin{equation}{section}

\definecolor{mygray}{rgb}{1, .5, .25}

\title{A New Class of de Sitter Vacua in Type IIB Large Volume Compactifications} 

\author[1]{Diego Gallego}
\author[2]{, M.C.~David Marsh}
\author[3]{, Bert Vercnocke}
\author[4]{and Timm Wrase}
\affiliation[1]{Escuela de F\'isica, Universidad Pedag\'ogica y Tecnol\'ogica de Colombia (UPTC),\\
	Avenida Central del Norte, Tunja, Colombia}
\affiliation[2]{Department of Applied Mathematics and Theoretical Physics, University of Cambridge, \\
	Cambridge, CB3 0WA, United Kingdom}
\affiliation[3]{Institute for Theoretical Physics, KU Leuven, Celestijnenlaan 200D, B-3001
	Leuven, Belgium}
\affiliation[4]{Institute for Theoretical Physics, TU Wien, Wiedner Hauptstr. 8-10, A-1040 Vienna, Austria}
\emailAdd{diego.gallego@uptc.edu.co}
\emailAdd{m.c.d.marsh@damtp.cam.ac.uk}
\emailAdd{bert.vercnocke@kuleuven.be}
\emailAdd{timm.wrase@tuwien.ac.at}

\abstract{
We construct a new class of metastable de Sitter vacua of flux compactifications of type IIB string theory. These solutions provide a natural extension of the `Large Volume Scenario' anti-de Sitter vacua, and can analogously be realised at parametrically large volume and weak string coupling, using standard ${\cal N}=1$ supergravity. For these new vacua, a positive vacuum energy is achieved from the inclusion of a small amount of flux-induced supersymmetry breaking in the complex structure and axio-dilaton sector, and no additional `uplift' contribution (e.g.~from anti-branes) is required. We show that the approximate no-scale structure of the effective theory strongly influences the spectrum of the stabilised moduli: one complex structure modulus remains significantly lighter than the supersymmetry breaking scale, and metastability requires only modest amounts of tuning. After discussing these general results, we provide a recipe for constructing de Sitter vacua on a given compactification manifold, and give an explicit example of a de Sitter vacuum for the compactification on the Calabi-Yau orientifold realised in $\mathbb{CP}^4_{11169}$. Finally, we note that these solutions have intriguing implications for phenomenology, predicting no superpartners in the spectrum below $\sim$50 TeV, and no WIMP dark matter. 
	}

\begin{document}
	
	\makeatletter
	\let\old@fpheader\@fpheader
	\renewcommand{\@fpheader}{
		\hfill 
	}
	\makeatother
	
	\maketitle
	
	\section{\label{sec:intro} Introduction}
	There are apparently many different possibilities to compactify string theory to four dimensions.  Most notably, generalised electromagnetic fluxes (e.g.~$F_3$ and $H_3$ for type IIB compactifications) can thread non-trivial cycles of the compactification geometry in a myriad of ways. For each viable compactification manifold, this results in a set of `flux compactifications', each with a distinct low-energy limit (for some reviews see \cite{Grana:2005jc,Douglas:2006es,Denef:2008wq}). The number of effective theories in this set grows exponentially  with certain topological parameters of the compactification manifold \cite{Douglas:2003um,Ashok:2003gk,Denef:2004ze,Denef:2004cf}.  However, while flux compactifications produce \emph{many} low-energy effective descriptions, it is not known whether \emph{any} of them  
	support solutions that are consistent with all experiments and observations. Consequently, it is not known if our own universe is included in this set of solutions, and if so, what (if any) the distinguishing characteristics of the physically interesting solutions are. 
	
	A key challenge in determining the properties of generic flux compactifications is that most vacua arise from the topologically richest and most complicated manifolds, obscuring the connection with observation. Deformation modes of the compact geometry appear in the low-energy theory as light `moduli' fields. Topologically interesting compactifications come with many moduli: typical type IIB scenarios have ${\cal O}(100)$ moduli fields \cite{Kreuzer:2000xy},  while for F-theory scenarios that number easily reaches ${\cal O}(10^5)$ \cite{Lynker:1998pb}. Those moduli do not only appear as exotic particles in the spectrum but also control important parameters in the low energy theory such as coupling constants. This brings the need for a controlled stabilisation, with non-zero masses and fixed vacuum expectation values. 
	
	The stabilised geometry, free of tachyons, corresponds to critical points of the effective potential that satisfy a perturbative (meta-)stability condition. For a configuration with $\partial_A  V = 0$, where $\partial_A = \frac{\partial}{
		\partial \phi^A}$ for the moduli fields $\phi^A$, this means that the Hessian matrix, 
	\be
	{\cal H} =
	\left(
	\begin{array}{c c}
		\partial^2_{A \bar B} V & \partial^2_{A B} V \\
		\partial^2_{\bar A \bar B} V & \partial^2_{\bar A  B} V
	\end{array} 
	\right) \, ,
	\ee 
	has only positive eigenvalues. 
	Stability is guaranteed for supersymmetric solutions. However, 
	the accelerated expansion of the universe and 
	the absence of observed superpartners to the Standard Model particles means that supersymmetry must be broken in solutions aiming to describe the real world. This motivates studying non-supersymmetric solutions of string compactifications. Ensuring 	a positive definite spectrum of the high-dimensional matrix ${\cal H}$ in solutions with broken supersymmetry is 
	a non-trivial task. 
	
	Moreover,  to address the accelerated expansion of the universe, 
	the 
	vacuum should be a 
	de Sitter solution with a positive cosmological constant \cite{Riess:1998cb,Perlmutter:1998np}. There are good reasons to believe that string compactifications allowing de Sitter solutions will be quite special and may share characteristic properties that could possibly lead to observational signatures. %
	One piece of evidence for this comes from theories  with many randomly interacting fields which can be studied by the means of random matrix theory (RMT). In such theories, metastable de Sitter vacua are exceptionally rare  \cite{Aazami:2005jf}; this conclusion remains unchanged even for typical critical points of supergravity theories with spontaneously broken supersymmetry \cite{Marsh:2011aa}. Hence, some type of non-random structure inherited from the string compactification appears to be required to explain the accelerated expansion of the universe.\footnote{%
	Recently,
several attempts have been made to	model the low-energy effective theories from string theory using classes of random functions, in particular Gaussian Random Fields (GRFs). These studies have found that, at least for trivial field space geometry and a particular choice of covariance function \cite{Bachlechner:2014rqa}, or upon ignoring the structures imposed by supergravity \cite{Bray:2007tf, Easther:2016ire}, metastable vacua  
are more common
 than the  random matrix theory argument suggests. 
 In the simplest models where $V$ is postulated to be a mean-zero Gaussian Random Field with a Gaussian covariance function, 
 the increase in the number of vacua can be understood to arise from a rigid shift in the eigenvalue spectrum  \cite{Bray:2007tf, Bachlechner:2014rqa}: for $V=0$, the GRF and RMT both predict a mean-zero Wigner semi-circle spectrum, but for $V>0$ the spectrum of the GRF potential  are shifted downwards and metastability is \emph{more} rare in the GRF than  the RMT analysis suggests. Conversely, for $V<0$ the spectrum is shifted to positive values and metastable vacua are more common. Hence, the GRF analysis suggests that the RMT estimate provides an \emph{upper bound} on the the frequency of metastable de Sitter vacua in random potentials. 
	} 
	
	The most studied class of metastable de Sitter solutions arising from string theory involve the `uplift' of a supersymmetric anti-de Sitter solution to positive vacuum energy by the means of a set of supersymmetry breaking anti-branes placed in a warped throat of the compactification manifold. By tuning the fluxes, many moduli can be stabilised supersymmetrically at a parametrically high scale, and are rather insensitive to the uplift. The low-energy effective field theory derived from dimensional reduction of this theory can be rephrased as an interesting version of ${\cal N}=1$ supergravity \cite{Ferrara:2014kva, Kallosh:2014wsa, Bergshoeff:2015jxa, Kallosh:2015nia, Aparicio:2015psl, Garcia-Etxebarria:2015lif, Dasgupta:2016prs, Vercnocke:2016fbt, Kallosh:2016aep, Aalsma:2017ulu}. Alternative proposals for constructing 
	Minkowski or de Sitter vacua
	either deform the theory, as in the so-called K\"ahler uplift \cite{Westphal:2006tn, Sumitomo:2013vla}, or extend its field content, by e.g.~the addition of open string fields and gauge dynamics \cite{Achucarro:2006zf, Dudas:2006vc, Dudas:2006gr, Abe:2006xp, Parameswaran:2007kf, Lalak:2007qd, Abe:2007yb, Brax:2007xq, Brax:2007fe, Dudas:2007nz, Gallego:2008sv, Covi:2008zu, Giveon:2009yu, Cicoli:2012fh, Cicoli:2013rwa, Rummel:2014raa, Kallosh:2014via, Cicoli:2015ylx}. However, these solutions may ultimately only correspond to a particular branch of a much wider set of metastable de Sitter solutions, and it is possible that more general and simpler solutions remain to be discovered. 
	
	The purpose of this paper is to construct a new branch of metastable de Sitter solutions of type IIB flux compactifications at large volume. These solutions are in some sense simpler than those obtained from anti-brane breaking: the de Sitter solutions presented in this paper require only the minimal set of ingredients being generated simply from spontaneous supersymmetry breaking in the moduli sector, including a comparatively small amount of supersymmetry breaking from the complex structure moduli and axio-dilaton sector. Consequently, all relevant physics should be captured by the standard ${\cal N}=1$ supergravity describing the low-energy limit of string compactifications. 
	
	The new solutions arise as a direct generalisation of those in \cite{Marsh:2014nla} (see also \cite{Kallosh:2014oja}) by including multiple K\"ahler moduli and thereby allowing for a larger compactification volume that in turn is a necessary condition for our construction. Similar ideas have been explored numerically before in particular examples by allowing relevant contributions to the supersymmetry breaking from the axio-dilaton and complex structure sector,
	either by 
	%
	 %
	%
	stabilising 
	the K\"ahler sector  as 
	in the KKLT scenario 
	\cite{Saltman:2004sn}, 
	or by non-geometric fluxes in STU-models \cite{Blaback:2015zra}, or by modelling the spectrum by random matrix theory \cite{Achucarro:2015kja}.
	
	
	The de Sitter and Minkowski vacua reported here allow for an analytic treatment 
	and 
	extends the vacua constructed in the so-called Large Volume Scenario (LVS) \cite{Balasubramanian:2005zx,Conlon:2005ki}.
	Notably, we will present explicit  de Sitter solutions, building on LVS, for compactifications on the Calabi-Yau orientifold realised as a hypersurface in $\mathbb{CP}_{11169}^4$. 
	
	A key benefit of these solutions is that they render almost all moduli metastable, without much tuning. This is achieved by only weakly breaking the leading order no-scale symmetry of the four-dimensional supergravity, and by utilising the lingering decoupling between the K\"ahler moduli sector and the complex structure moduli and the axio-dilaton.
	
	These results deepen our understanding of the vacuum structure of the effective theories arising from flux compactifications and  may have interesting applications for cosmological and particle physics models in string theory. They may also serve as a rather explicit testing ground for conjectures about non-supersymmetric vacua in string theory \cite{Ooguri:2016pdq, Freivogel:2016qwc, Danielsson:2016mtx}.
	

	\section{Review of type IIB flux compactifications}
	In this section, we review the structure of supergravity theories in four dimensions descending from the low-energy limit of IIB string theory on a Calabi-Yau orientifold with RR and NSNS fluxes.\footnote{We anticipate that our analysis and results extend straight-forwardly to F-theory compactifications on CY$_4$-manifolds.} In particular: in section \ref{sec:Kahler} we review the K\"ahler geometry of the moduli space; in section \ref{sec:flux} we discuss the flux induced superpotential and possible non-perturbative corrections; and in section \ref{sec:decomp} we detail how physical quantities relevant for the four-dimensional effective theory are captured by the various components of the flux vector. 
	
	We consider the four-dimensional spectrum of a compactification on the orientifold $\tilde M_3$ of the Calabi-Yau three-fold $M_3$. 
	The low-energy degrees of freedom include the axio-dilaton $S = C_0 + \rmi e^{-\phi}$,  the complex structure moduli $u^i$, where $i=1,\ldots,h^{1,2}_-(\tilde M_3)$, and the K\"ahler moduli $T^a$, where $a=1,\ldots, h^{1,1}_+( \tilde M_3)$ \cite{Grimm:2004uq}. The corresponding low-energy theory can often be described by four-dimensional ${\cal N}=1$ supergravity in which the moduli furnish the scalar components of chiral multiplets. Our convention for indices is as follows: $A,\, B$ etcetera run over all moduli fields, $I,\, J$ run over all complex structure moduli and the axio-dilaton, $i,\, j$ run over complex structure moduli only,  $a,\, b$ run over K\"ahler moduli only, and finally, $s$ runs over `blow-up' K\"ahler moduli:
	\be
	X^A \equiv (S,u^i,T^a)\,,\qquad X^I \equiv (S,u^i)\,,\qquad T^a \equiv (T^{\rm big}, T^s)\,.
	\ee

	\subsection{Moduli space geometry}\label{sec:Kahler}
	The kinematics of the moduli fields are governed by a real K\"ahler potential, $K$,  	which at  sufficiently large volume and weak string coupling can be written as the sum of three contributions,
	\be
	\label{eq:Ktot}
	K=\Ks+ \Kcs + \Kk \, .
	\ee
	We will throughout this paper use the shorthands $K_i = \partial_{u^i} K$, $K_a  = \partial_{T^a} K$,  $K_S = \partial_{S} K$, etc, and we will also use the notation $\tilde K = \Ks +\Kcs$. The explicit expression for \eqref{eq:Ktot} is determined by the geometry and topology of the compactification manifold, as we now briefly review.
	
	

	\subsubsection{Axio-dilaton and complex structure deformations}
	%
	The leading-order K\"ahler potential for the axio-dilaton $S$ is given by,
	\be
	\Ks = -\ln \ls -\rmi (S-\bar S) \rs \, .
	\ee
	Because of the $SL(2,\mathbb{Z})$ S-duality, the axio-dilaton $S$ can be restricted to the fundamental domain of the torus, $\{ S \in \mathbb{C},~|{\rm Re}\,S| < \frac 12,~ |S|\geq1,~{\rm Im}\,(S) > 0\}$.

	The complex structure moduli space is conveniently parametrised by means of projective coordinates $z^I, I = 0,1,\ldots h^{1,2}$, which correspond to the periods of the holomorphic three-form $\Omega$,
	\be
	\vec{\Pi} =
	\left(
	\begin{array}{c}
		\int_{A^I} \Omega \\
		\int_{B_I} \Omega
	\end{array}
	\right)
	\equiv 
	\left(
	\begin{array}{c}
		z^I \\
		{\cal G}_I
	\end{array}
	\right) \, ,
	\label{eq:Period0}
	\ee
	where we chose a canonical symplectic basis $(A^I,B_I)$ of the the third homology group of $H^3(M_3)$ and $(\alpha_I,\beta^I)$ denote the dual cohomology group basis
	\bea
	\int_{M_3}\alpha_I \wedge \beta^J = - \int _{M_3} \beta^J \wedge \alpha_I = \delta_I^J\,,\qquad \int_{M_3}\alpha_I \wedge \alpha_J = \int_{M_3} \beta^I \wedge \beta^J = 0\,.
	\eea
	The periods  ${\cal G}_I$ satisfy the condition
	$
	2 {\cal G}_I = \del_I (z^J {\cal G}_J)
	$, thus form the gradient of a function that is homogenous of degree two: ${\cal G}_I = \partial_I {\cal G}$. 
	
	The set of inhomogeneous coordinates on the complex structure moduli space is conventionally chosen as,
	\be
	u^i = z^i/z^0\,,\qquad i = 1, \ldots, h^{1,2}\,.
	\ee
	Upon setting $z^0 =1$, the period vector is given by.
	\be
	\Pv = 
	\left(
	\begin{array}{c}
		1 \\
		u^i \\
		2 {\cal F}- u^j {\cal F}_j \\
		{\cal F}_i
	\end{array}
	\right) \, ,
	\label{eq:Period}
	\ee
	with the prepotential ${\cal F} = \G/(z_0)^2$. 
	
An ${\cal N}=1$ supergravity theory in $d=4$, at low energies, is obtained by working not in $M_3$ but on its orientifold image $\tilde M_3$, where only the involution-odd complex structure moduli are kept in the chiral spectrum. Since we will not be concerned with the details of this involution, we will simply restrict $u^i$ to run over $i=1,\ldots, h^{1,2}_-$.

	The complex structure-dependent  contribution to  the K\"ahler potential is given by,
	\bea
	K_{\rm c.s.} &=& - \ln \left(\rmi \int_{M_3} \Omega \wedge \bar \Omega \right) = 
	-\ln \left( \rmi \vec{\Pi}^{\dagger}\, \Sigma\, \vec{\Pi} \right)
	\, ,
	\eea
	where $\Sigma$ denotes the symplectic metric,
	\be
	\Sigma = 
	\left(
	\begin{array}{c c}
		0 & \mathds{1} \\
		- \mathds{1} & 0
	\end{array}
	\right) \, .
	\ee

	\subsubsection{K\"ahler deformations}
	
	
	The imaginary component of the K\"ahler moduli fields $T^a$ are given by the  four-cycle volumes, $\tau^a$,  which can be defined from the Calabi-Yau manifold volume, $\V$. 
	The volume is a homogeneous function of degree 3 in the two-cycle volumes
	 $t_a$, and  the four-cycle volumes are defined by $\tau^a = \V_{t_a}$. It then follows that,
	\be
	\label{eq:32}
	t_a \tau^a = 3 \V \, . 
	\ee 
	
	Our focus in this paper is on de Sitter vacua in the  Large Volume Scenario \cite{Balasubramanian:2005zx,Conlon:2005ki}, for which it is convenient to consider compactifications volumes of the (strong) `Swiss cheese' type,
	\be
	\V = (\eta_{\rm big} \tau^{\rm big})^{\frac32} - \sum_{s=1}^{N_{\rm small}} (\eta_s  \tau^s)^{\frac32}\,,
	\label{eq:SC}
	\ee
	where $\eta_a \in \mathbb{R}$. We have also split the index $a=1,\ldots, h^{1,1}_+$ into one `big' cycle  and $s=1,\ldots, N_{\rm small}$ `small' cycles. Here $N_{\rm small}=h^{1,1}_+-1$ is the number of blow-up cycles in the compactification geometry.
	
	The K\"ahler potential for the K\"ahler moduli is given by \cite{Becker:2002nn}, 
	\be
	\label{eq:KK}
	\Kk= -2 \ln \ls \V + \frac{\xi}{2} \lp -\rmi \frac{S-\bar S}{2} \rp^{\frac{3}{2}} \rs\, .
	\ee
	The contribution multiplying $\xi = -\frac{\chi(CY_3) \zeta(3)}{2 (2\pi)^3}$ arises from $(\alpha')^3$ corrections in the ten-dimensional theory. At sufficiently large volume, this term provides the dominant correction to the resulting scalar potential \cite{Berg:2007wt, Cicoli:2007xp}.  The $\alpha'$-correction induces non-vanishing cross-terms between the axio-dilaton and K\"ahler moduli components of the metric. The relevant components of the inverse metric are given by \cite{Blumenhagen:2009gk},
	\bea
	K^{a \bar b}_K &=& -2 \left( {\cal V} + \frac{\hat \xi}{2} \right) \left(\frac{\partial^2 {\cal V}}{\partial \tau^a \partial \tau^b }\right)^{-1} + \tau^a \tau^b \frac{4 {\cal V} - \hat \xi}{{\cal V}- \hat \xi } \, , \\
	K^{a \bar S}_K &=& i  \frac{3}{2} ( S- \bar S) \tau^a \frac{\hat \xi}{{\cal V} - \hat \xi} \, , \\
	K^{S \bar S}_K &=& -\frac{(S-\bar S)^2}{4} \frac{4 {\cal V} - \hat \xi}{{\cal V}- \hat \xi} \, ,
	\eea  
	where $\hat \xi = \xi (-\rmi (S-\bar S)/2)^{3/2}$. 

	\subsection{Flux compactifications} \label{sec:flux}
	
	We are interested in compactifications in which 
	integrally quantised RR ($F_3$) and NSNS ($H_3$) fluxes thread some non-trivial three-cycles of $M_3$, 
	\be
	\frac{1}{(2\pi)^2 \alpha'} \int_{A^I, B_I} F_3 = \vec{N}_{\rm RR} \in \mathbb{Z}^{2(h^{1,2}_-+1)}\, ,~~~\frac{1}{(2\pi)^2 \alpha'} \int_{A^I, B_I} H_3 = \vec{N}_{\rm NSNS} \in \mathbb{Z}^{2(h^{1,2}_-+1)}\, .
	\ee
	It is conventional to introduce the complex three-form flux $G_3 = F_3 - S H_3$, and convenient to define the complexified flux vector  as,
	\be
	\vec N 
	= - \left(
	\begin{array}{c}
		\int_{A^I} G_3 \\
		\int_{B_I} G_3
	\end{array}
	\right) \, .
	\ee 
	The fluxes contribute to the D3-charge tadpole by, 
	\be
	Q_{\rm flux} = \frac{1}{(2\pi)^4 (\alpha')^2} \int_{M_3} H_3 \wedge F_3 
	= 
	\frac{1}{(2\pi)^4 (\alpha')^2} 
	\vec{h}^T \Sigma \vec{f} 
	=
	- \frac{1}{(2\pi)^4 (\alpha')^2} 
	K_{S}
	\vec N^{\dagger}\, \Sigma\, \vec N
	\, ,
	\label{eq:Qflux}
	\ee 
	where,
	\be
	\vec{f} = \left(
	\begin{array}{c}
		\int_A F_3 \\
		\int_B F_3 
	\end{array}
	\right) \, , 
	~~{\rm and}~~
	\vec{h} = \left(
	\begin{array}{c}
		\int_A H_3 \\
		\int_B H_3 
	\end{array}
	\right) \, .
	\ee
	The relation between $\vec N$ and $\vec f$ and $\vec h$ is then clearly,
	\be
	{\rm Re}(\vec N) = - \vec f_3 + {\rm Re}(S) \vec h_3 \, , 
	~~{\rm and}~~
	{\rm Im}(\vec N) = {\rm Im}(S) \vec h_3 \, .
	\ee
	
	Requiring that the total sum of all D3 charges vanishes in the internal space leads to a joint condition on the D3-brane content, the fluxes, and the D7-brane and O-plane configuration, 
	\be
	Q_{\rm flux} + N_{\rm D3} = \frac{\chi}{24} \, ,
	\ee
	where $N_{\rm D3}$ denotes the net number of D3-branes, and $\chi$ is the Euler characteristic of the Calabi-Yau fourfold that corresponds to the F-theory lift of our type IIB compactification.  Fluxes that preserve  supersymmetry contribute positively to this tadpole condition.
	
	The fluxes induce a complex structure and axio-dilaton dependent energy density that in the four-dimensional theory is captured by  the flux induced superpotential \cite{Gukov:1999ya},
	\be
	W_0 =
	\int_{M_3}  G_3 \wedge \Omega =  - \vec{N}^T\, \Sigma\, \vec{\Pi} 
	\, .
	\label{eq:GVW}
	\ee 
	This superpotential is exact to all orders in perturbation theory, and does not receive $\alpha'$-corrections. However, non-perturbative contributions from Euclidean D3-branes and gaugino condensation on stacks of D7-branes induce  additional contributions: 
	\bea\label{eq:W}
	W&=& W_0 + W_{\rm np} = W_0 + \sum_{s=1}^{N_{small}} A_s\, e^{\rmi a_s T^s}\,,
	\eea
	where we have specialised to compactifications of `Swiss-cheese' form \eqref{eq:SC}, and included non-perturbative corrections for all blow-up moduli $T^s$,  but -- anticipating solutions with large volume --  neglected such a correction for the single `big cycle' $T^{\rm big}$. The factor $a_s$ is $a_s =2\pi$ for Euclidean D3-branes and $a_s = 2\pi/N_{\rm D7}$ for a stack of $N_{\rm D7}$ branes. The prefactors $A_s$ depend on the axio-dilaton and the complex structure moduli. The moduli dependence  can in principle be determined from the Pfaffian of one-loop corrections to the instanton action, however such a computation is very difficult. 
As we show below, the particular functional forms of the prefactors $A_s$ are generically not important for our analysis and we can consistently treat them as constants.
	
	In this paper we denote by $D_A$ the K\"ahler and geometrically covariant derivative, and we use the following notation for covariant derivatives of the superpotential,  
	\bea
	F_A &=& D_A W = \partial_A W + K_A W \, , \nonumber \\
	Z_{AB} &=& D_A F_B = \partial_A F_B + K_A F_B - \Gamma_{AB}^C F_C \, ,
	\nonumber \\
	U_{ABC} &=& D_A Z_{BC} =D_B Z_{AC} =D_C Z_{BA}
	\, .
	\label{eq:FZU}
	\eea
		The supergravity F-term scalar potential is denoted by,
	\be
	V=e^{K}\Big(F_A \bar  F^A-3|W|^{2}\Big)\,,
	\ee
	where indices are raised with the inverse {\K} metric $K^{A\bar B}$.
The gravitino mass is given by,
\be
m_{3/2} = e^{K/2} |W| \, .
\ee

	\subsection{Decomposition of flux vector} \label{sec:decomp}
	The three-form $G_3$ is conveniently decomposed by Hodge type, and it is well known that supersymmetric fluxes are  of Hodge type (2,1) and primitive \cite{Grana:2000jj, Gubser:2000vg, Grana:2001xn}. This condition is manifest in the low-energy effective description in that, at the complex structure moduli space location of the vacuum,
	the flux vector $\vec N$ only has non-vanishing components along the $D_i \Pv$ directions. The de Sitter solutions that we construct will involve non-supersymmetric flux vectors that in addition to the (2,1) components have a non-vanishing (0,3) piece (corresponding to $W_0 \neq 0$), and comparatively smaller contributions along the (1,2) and (3,0) directions (corresponding to $F_i,~F_S \neq 0$). 
	
	To see the relation between physical quantities and fluxes explicitly, we first recall that $\{\Omega, \bar \Omega, D_i \Omega, \bar D_{\ib} \bar \Omega\}$ form a basis of three-forms for generic values of the moduli \cite{Candelas:1990pi}.  
	Correspondingly, $\{\Pv, \Pv^*, D_i \Pv, \bar D_{\ib} \Pv^* \}$ form an orthogonal basis with respect to the symplectic inner product \cite{Denef:2004ze,Brodie:2015kza, Marsh:2015zoa}, 
	\bea
	\Pv^T\, \Sigma\, \Pv^* &=& + \rmi e^{-\Kcs} \, , \\
	D_i \Pv^T\, \Sigma\, \bar D_{\jb} \Pv^* &=& -\rmi K_{i \jb} e^{-\Kcs} \, ,\label{eq:DPiDPibar} \\
	 \Pv^T\, \Sigma\, D_{i} \Pv &=& \Pv^T\, \Sigma\, \bar  D_{\ib} \Pv^* =D_{i} \Pv^T\, \Sigma\, D_{j} \Pv = 0 \, .
	\eea
	Hence, any vector $\vec V$ in the `period/charge' vector space can be expanded as follows:
	\be
	\vec V = \rmi e^{\Kcs} \left(
	(\Pv^\dagger \Sigma \vec V) \Pv - (\Pv^T \Sigma \vec V) \Pv^* - (K^{\jb l} \bar D_{\jb} \Pv^\dagger \Sigma \vec V) D_{l} \Pv +
	(K^{j \bar  l}  D_{ j} \Pv^T \Sigma \vec V) \bar D_{\bar l} \Pv^* 
	\right) \, , \nonumber 
	\ee 
	for $\Pv$ evaluated at a generic point in the complex structure moduli space. 
	For the flux vector $\vec N$, the expansion coefficients have direct physical interpretations,
	%
	\be
	\vec N = \rmi e^{\Kcs} \left(
	\frac{\bar F_{\bar S}}{K_{\bar S}} \Pv-W_0 \Pv^* - \frac{\bZ^i_{~\bar S}}{K_{\bar S}} D_i \Pv + F^{\ib} \bar D_{\ib} \Pv^*
	\right) \, .
	\label{eq:NPi}
	\ee
	which follows straightforwardly from the superpotential and by using $D_{S} \vec N = K_{S} \vec{N}^*$.

	We  note that $D_i D_j \Omega$ is purely (1,2) and can be expanded using only the three-forms $\bar D_{\ib} \bar \Omega$.
Therefore the component $Z_{ij} = D_i D_j W$ of the tensor are not free variables, but given in terms of the fluxes $\bar Z_{\ib \bar S}$ as,
	\be
	Z_{ij} = +\rmi e^{\Kcs} \kappa_{ij}^{~~\bar k} \vec N\, \Sigma\, \bar D_{\bar k} \Pv^* 
	= - \frac{\rmi}{K_{\bar S}} e^{\Kcs}\kappa_{ij}^{~~\bar k} \bZ_{\bar k \bar S}
	\, ,
	\label{eq:Zij}
	\ee
	where $\kappa_{ijk}$ denotes the (in general field-dependent) `Yukawa couplings', $\kappa_{ijk} = \int_{M_3} \Omega \wedge \partial^3_{ijk} \Omega$. 
	The flux tadpole contribution may now be expressed as,
	\be
	(2\pi)^4 (\alpha')^2 \, Q_{\rm flux} = -\rmi K_{S} e^{\Kcs} \left(
	|W_0|^2 + \frac{1}{|K_{S}|^2} Z_{S l} \bZ^l_{~\bar S} - F_i \bar F^i - \Big| \frac{F_{S}}{K_{S}}\Big|^2
	\right) \, . 
	\label{eq:tadpole}
	\ee
	We will find the decomposition of the flux vector \eqref{eq:NPi} very useful in constructing explicit  de Sitter vacua in section \ref{sec:example}.
	
	\section{No-scale symmetry, slightly broken 
	}\label{sec:review1stpaper}
	In this section we review the implications of no-scale symmetry for the low-energy effective theory from string compactifications. Our particular focus will be on the metastable de Sitter vacua constructed in \cite{Marsh:2014nla}, for which the slightly broken no-scale symmetry bestows  favourable metastability properties. 
	%
	\subsection{No-scale symmetry}
	Four-dimensional ${\cal N}=1$ supergravity theories are said to be `no-scale' \cite{Cremmer:1983bf,Ellis:1983ei,Ellis:1984bm} if  a subset of the fields, here denoted $T^a$, 
	\begin{itemize}
	\item[{\it i)}] have no K\"ahler potential  cross-couplings with the other fields, here denoted $X^I$, so that $K = K_1(X^I) + K_2(T^a)$, 
	\item[{\it ii)}] do not appear in the superpotential,  $W = W_0 (X^I) $,\footnote{A weaker form of `generalised no-scale condition' takes $W(X^I, T^a)$ but postulates that   the first three derivatives of $W$ with respect to the fields $T^a$ vanish  at the  critical point \cite{Danielsson:2015rca}. } 
	\item[{\it iii)}] have a field-space geometry satisfying the no-scale condition,
	\be
	K_a K^{a \bar b} K_{\bar b} = 3 \, .
	\label{eq:noscaleK}
	\ee 
	\end{itemize}
	Such theories can admit solutions to the critical point equation, $\partial_A V = 0$, with $F_I= 0$ and $F_a = K_a W$, so that,
	 	\be
	V = e^K \left(F_A \bar F^A - 3 |W|^2 \right) = e^K F_I \bar F^I  = 0 \, .
	\label{eq:noscale}
	\ee
	In these non-supersymmetric Minkowski solutions, the fields $X^I$ are 
	stabilised while the fields $T^a$ remain unfixed. Specifically, for these solutions $\partial^2_{T \bar T} V = \partial^2_{T  T} V = \partial^2_{i \bar T} V = \partial^2_{i  T} V = 0$.
	The Hessian matrix for the fields $X^I$ is, in the notation of equation \eqref{eq:FZU}, given by,
	\be
	{\cal H}
	= \left(
	\begin{array}{c c}
		\partial^2_{I\bar J} V& \partial^2_{IJ} V  \\
		\partial^2_{\bar I  \bar J } V & \partial^2_{\bar I  J} V
	\end{array}
	\right)=
	e^K\left(\begin{array}{cc} (Z \bZ)_{I \bar J}+K_{I\Jb} |W|^2 & 2\bW Z_{IJ } \\
		2W\bar Z_{\Ib\Jb } & (\bZ Z)_{\bar I J}+K_{\Ib J} |W|^2\end{array}\right)  \,.
	\ee
For canonically normalised fields\footnote{We call fields $\tilde{\phi}^I$ canonically normalised, if they satisfy $K_{I\bar J} \partial_\mu \phi^I \partial^\mu \bar{\phi}^{\bar J} = \delta_{I\bar J} \partial_\mu \tilde{\phi}^I \partial^\mu \bar{\tilde{\phi}}^{\bar J}$, and we call the eigenvalues of the Hessian the masses squared. } it has the eigenvalues \cite{Marsh:2014nla},
	\be
	m^2_{I\pm} = e^K \left( \lambda_I \pm |W|\right)^2 \, ,
	\label{eq:msq0}
	\ee 
	where $\lambda_I^2$ is an eigenvalue of $(Z \bZ)_{I \bar J} \equiv Z_{IK} \bZ^K_{~\bar J}$. The semi-positive-definiteness of the eigenvalues of the Hessian follows from the convexity of the potential \eqref{eq:noscale}, and it has important consequences for the metastability of de Sitter vacua constructed from theories with slightly broken no-scale symmetry.

	To lowest order in the $g_s$ and $\alpha'$ expansions of type IIB string theory, the K\"ahler potential of the K\"ahler moduli sector is given by  $K_K= - 2 \ln {\cal V}$, cf. \eqref{eq:KK}, and  there are no cross-couplings between K\"ahler moduli and the complex structure/axio-dilaton moduli sector, i.e.~$X^I = (u^i, S)$. 
	The geometric condition that the volume is a homogeneous function of degree $3/2$ in the four-cycle volumes, cf. equation \eqref{eq:32}, 
	implies that   $K^{a \bar b} K_a K_{\bar b} = 3$.
	Moreover, to this order, the superpotential is simply given by the K\"ahler moduli independent flux-induced contribution \eqref{eq:GVW}, and  the K\"ahler moduli enjoy a leading order no-scale symmetry. Consequently, the complex structure moduli and the axio-dilaton have squared masses given by \eqref{eq:msq0}, once canonically normalised.

	\subsection{Approximately no-scale de Sitter vacua}
	\label{sec:approxnoscalereview}
		
	In type IIB compactifications at low energies the no-scale symmetry is only approximate, broken by both string loop and  sigma model corrections.
	For example, the leading $(\alpha')^3$ correction captured by the K\"ahler potential \eqref{eq:KK} implies that, 
	\be
	K_a = -2 \frac{\V_a}{\V} \left(1 - \frac{\hat \xi}{2 \V} \right) \, ,
	\ee
	so that the $\alpha'$-corrections break the no-scale condition by a small, volume suppressed, amount:
	\be\label{eq:no-scaleapprox}
	K^{a \bar b} K_a K_{\bar b}= 3 + \frac{3}{4} \frac{\hat \xi}{\V} \, .
	\ee 
	This small breaking is of critical importance to the AdS vacua of the Large Volume Scenario (as we review in section \ref{sec:LVSreview}), and will become equally important  for our de Sitter vacua constructed in section \ref{sec:dSvac}.  For the remainder of this section  however, we take $\hat \xi \to 0$ and consider a different type of no-scale breaking that arises from a postulated, small explicit \emph{superpotential} deformation of the theory.

	More precisely, following \cite{Marsh:2014nla}, we specialise to the single K\"ahler modulus case and  take,
	\be
	W(T, X^I) = W_0(X^I) + \delta W(T, X^I) \, .
	\ee
	The superpotential deformation $\delta W$ (and any derivative of it) is assumed to be small compared to $W_0$: $|\delta W/W_0| \ll 1$. 
	The F-terms are now given by,
	\be
	F_T = K_T W + \delta W_T \,, ~{\rm and} ~~~F_I = \epsilon W\, f_I \, ,
	\ee
	where $\epsilon \ll 1$ and $f_I$ is a unit vector. A consistent solution of the critical point equation $\partial_T V= 0$ takes $\delta W$ (and any derivatives of it) to be of ${\cal O}(\epsilon^2 W)$. The critical point equations for the remaining fields, $\partial_I V = 0$, imply that,
	\be \label{eq:cpcs}
	Z_{IJ} \bar F^J = - \bW F_I \, ,
	\ee
	up to corrections of ${\cal O}(\epsilon^2)$. Note that $Z_{IJ}$ is a complex symmetric tensor and hence not unitarily diagonalisable. However, equation \eqref{eq:cpcs} may be recast as an eigenvalue equation for the Hermitian matrix $(Z\bZ)_{I \bar J}$, the eigenvalue $|W|^2$ and the eigenvector $F_I$:
	\begin{equation}\label{eq:Z2EigenVal}
	(Z\bar Z)_{I}^{~J}F_{J}=|W|^2 F_I \, .
	\end{equation}
	An immediate and important implication of equation \eqref{eq:Z2EigenVal} is that one of the eigenvalues $\lambda_I^2$ of $(Z\bZ)$ will be precisely equal to $|W|^2$, and hence, according to equation \eqref{eq:msq0}, there is one real degree of freedom in the complex structure and axio-dilaton sector that is massless to leading order:
	\be
	m^2_{1\pm} = \left\{
	\begin{array}{l r}
		2e^K |W|^2  + {\cal O}(\epsilon |W|^2)& , \\
		{\cal O}(\epsilon |W|^2) & .
	\end{array}
	\right.
	\ee
	To order ${\cal O}(\epsilon^0)$, the  massless spectrum  of these solutions contains a total of three modes: the real and imaginary parts of $T$ and $X^{1-}$. 
	
	To linear order in $\epsilon$, the relevant part of the Hessian matrix is given by,
	\be
	{\cal H}
	= \left(
	\begin{array}{c c}
		\partial^2_{I\bar J} V& \partial^2_{IJ} V  \\
		\partial^2_{\bar I  \bar J } V & \partial^2_{\bar I  J} V
	\end{array}
	\right)=
	e^K\left(\begin{array}{cc} (Z \bZ)_{I \bar J}+K_{I\Jb } |W|^2~& ~2\bW Z_{IJ } + U_{IJK }\bar F^K\\
		2W\bar Z_{\Ib\Jb }+   \bar U_{\Ib\Jb \bar K }F^{\bar K} ~& ~(\bZ Z)_{\bar I J}+K_{\Ib J} |W|^2\end{array}\right) \, .
	\label{eq:HessLin}
	\ee
	The field $X^{1-}$ is then lifted at ${\cal O}(\epsilon)$ with the squared mass,
	\be \label{eq:lightmodemass}
	m^2_{1-} = e^K {\rm Re}\left(U_{IJK} \bar f^I\bar f^J\bar f^K \bW \right) \epsilon \, ,
	\ee 
	in the generic case of $U_{IJK} \bar f^I\bar f^J\bar f^K \sim {\cal O}(1)$. This squared mass can be made positive by a moderate tuning of the phase of the superpotential.

	In sum, the full spectrum of these solutions then consists of three sets of fields:
	\begin{itemize}
		\item All $X^I$, except for $X^{1-}$, receive positive squared masses at ${\cal O}\left( {\rm max} (e^K\lambda_I^2, e^K |W|^2 \right)$, and are not destabilised by the inclusion of small amounts of supersymmetry breaking in the complex structure and axio-dilaton sector. 
		\item The field $X^{1-}$  is lighter than the other fields in this sector, but can be stabilised at ${\cal O}(F_I/W)\sim\epsilon$ by a small amount of tuning. 
		\item  The real and imaginary parts of $T$ are lifted at ${\cal O}(\epsilon^2)$ and the corresponding eigenvalues of the Hessian matrix are given by,
		\be
		m^2_{T\pm} =\left( - \frac{4K_T}{3} {\rm Re}\left(\bW \delta W_T \right) \pm | K^T \bW \delta W_{TTT} - \frac{4 K_T}{3} {\rm Re}(\bW \delta W_T)|\right)e^K
		\, .
		\ee  
		As shown in \cite{Marsh:2014nla}, both these eigenvalues can be rendered positive. However, in the case of a single K\"ahler modulus, the tuning is somewhat restrictive.  For example, in the case of $\delta W$ arising from a single non-perturbative effect, $\delta W = A(X^i) {\rm exp}\left( \rmi a T\right)$, positivity of $m^2_{T-}$ requires $a \tau <\sqrt{2}$, while de Sitter minima have $a\tau >1$. These conditions restrict the values of $\epsilon$ that are realisable, given the vacuum expectation values of $A$ and $W_0$: $1.47 |A|/|W_0| < \epsilon^2 < 1.57 |A|/|W_0|$. Hence, in this case $\epsilon \ll 1$ requires $|A| \ll W_0$.
	\end{itemize} 
	
	
	In this paper, we will go beyond the discussion of \cite{Marsh:2014nla} by including no-scale breaking effects in both the superpotential and the K\"ahler potential, and we will consider compactifications with an arbitrary number of moduli. In this arguably more interesting case, we will see that the conditions for metastable de Sitter vacua can be greatly relaxed.

	\section{de Sitter vacua at large volume}
	\label{sec:dSvac}
	\label{sec:dS}
	In this section, we describe how the prescription of reference \cite{Marsh:2014nla} can be generalised to produce metastable de Sitter vacua at exponentially large volume, thereby extending the non-supersymmetric AdS vacua of \cite{Balasubramanian:2005zx}.  
	
	\subsection{Anti-de Sitter  vacua in the Large Volume Scenario}
	\label{sec:LVSreview}
	We briefly recall the  prescription of the Large Volume Scenario (LVS) for constructing non-supersymmetric AdS  vacua \cite{Balasubramanian:2005zx}.  Starting from the $\alpha'$-corrected K\"ahler potential 
	 \eqref{eq:Ktot} and the flux-induced superpotential \eqref{eq:W}, the complex structure and axio-dilaton are stabilised by requiring that the three-form flux background is supersymmetric, i.e. $F_I = 0$. To zeroth order in $\alpha'$, the K\"ahler moduli sector is no-scale, and the squared masses of the complex structure moduli and the axio-dilaton are given by equation \eqref{eq:msq0}. Under these circumstances \eqref{eq:Z2EigenVal} does not imply  $\lambda_I = |W|$, and generically, the complex structure moduli and the axio-dilaton are stabilised at the scale ${\cal O}\left( {\rm max}( e^K \lambda_I^2, e^K |W|^2)\right)$ for each $\lambda_I$. These moduli are then generically decoupled from the details of the K\"ahler moduli stabilisation \cite{Gallego:2011jm,Gallego:2013qe, Sousa:2014qza}.
	
	Secondly, the axions of the `small' cycles, ${\rm Re}(T^s)$, are lifted by contributions to the potential of the form $e^K \left( W_{np} \bW + {\rm c.c.} \right)$. Minimising the resulting cosine potentials results in a scalar potential only involving the  overall volume $\V$ and the `small' K\"ahler moduli, $\tau^s$. The leading contributions to the scalar potential in an expansion in powers of the inverse volume are,
	\bea\label{eq:LVSpotential}
		V_{\rm LVS} &=& e^{\tilde K} \left[
	\frac{3 \xi ({\rm Im}(S))^{3/2}}{4\V^3 } |W_0|^2 
	- 
 \sum_{s}^{N_{\rm small}} 	\frac{4  a_s \tau^s |A_s W_0|}{\V^2}   e^{-a_s \tau^s} 
	+ \sum_{s}^{N_{\rm small}} \frac{8\sqrt{\tau^s}}{3\eta^{3/2}_s\V} a_s^2 |A_s|^2 e^{-2a_s \tau^s} 
	\right],
	\nonumber
	\\
	&\equiv&
	\frac{ a}{\V^3} - \frac{ \sum_{s=1}^{ N_{\rm small}} b_s x_s e^{-x_s}  }{\V^2} + \frac{ \sum_{s=1}^{ N_{\rm small}} c_s \sqrt{x_s} e^{-2 x_s}  }{  \V}		
	\, ,
	\eea
		where we have introduced the notation $x_s = a_s \tau^s$. 
	To consistently neglect higher order instanton corrections, we only consider solutions with $x_s> 1$ for all $s$.  
	
	The critical point equations $\partial_{x_s} V = 0$ are solved by,
	\be
	\Vmin = 2 \frac{ b_s}{ c_s} e^{x_s} \sqrt{x_s} \left( \frac{x_s-1}{4x_s-1}\right)\approx \frac{ b_s}{2 c_s} e^{x_s} \sqrt{x_s}
	\, ,
	\label{eq:taussol}
	\ee
	for each $x_s$. In the last step we approximate $x_s \gg 1$.  Using the above in $\partial_\V V=0$, implies that,
	\be
	a = 4 \sum_{s=1}^{N_{small}}\frac{ b_s^2}{c_s}x_s^{5/2}\frac{ (x_s-1) }{ (4 x_s-1)^2}\approx \frac 14 \sum_{s=1}^{N_{small}}\frac{b_s^2}{c_s}x_s^{3/2} \, .
	\label{eq:Vsol}
	\ee
	This critical point is the minimum of the potential for $\V$. Away from the critical point the potential approaches zero from below in the limit $x_s \to \ln {\cal V}$ and $\V \to \infty$:
	\be
	V_{\rm LVS} \sim -\frac{\ln \V}{\V^3} \to 0 \, .
	\ee 
	Consequently, the minimum is a non-supersymmetric AdS vacuum. The leading-order vacuum energy is given by,
	\be
	V_{\rm LVS}|_{\rm min} = 
-	\frac{2}{\Vmin^3}\left(\sum_s\frac{b_s^2(x_s-1)x_s^{3/2}}{c_s(4x_s-1)^2}\right) \approx-  \frac 1 {8\Vmin^3} \sum_s\frac{b_s^2}{c_s} \sqrt{x_s}\,.
	\ee

	\subsection{Large volume extrema}
	
	We now 	generalise the prescription of the Large Volume Scenario  to construct  de Sitter vacua with an exponentially large volume.  
	The key idea is, just as discussed in section \ref{sec:approxnoscalereview}, to consistently include  a small amount of flux-induced supersymmetry breaking in the complex structure and axio-dilaton sector, as parametrised by the F-terms, 
	\begin{equation}
	 F_I = \epsilon W f_I\,,\label{eq:FI_LVS}
	\end{equation}
	with $\epsilon \ll 1$ a small parameter and $f_I$ a unit vector.
	Our construction generalises that of \cite{Marsh:2014nla} by
	including  multiple K\"ahler moduli and additional no-scale breaking from $\alpha'$-corrections to the K\"ahler potential. 
	
	An immediate concern is that non-vanishing F-terms for the complex structure and the axio-dilaton 
	potentially destabilise  the potential and cause decompactification. The additional F-terms, $ F_I \bar F^I = F_i \bar F^i + F_S \bar F^S$, contribute to the potential like, 
	\bea
	V &=&	V_{\rm LVS} 	+  \frac{e^{\tilde K} F_I \bar F^I}{\V^2} \, .
	\label{eq:Vmodified}
	\eea
	Since $V_{\rm LVS} \sim {\cal V}^{-3}$, we see that generic amounts of supersymmetry breaking, $F_I \bar F^I \sim |W|^2$,  source a dominant run-away potential for the volume.\footnote{
	The existence of run-away directions in the moduli space has since long been identified as a critical aspect of string compactifications \cite{Dine:1985he}. 
	A detailed analysis of the $F_I\bF^I \sim |W|^2$	 case is intricate: for comparatively small volumes and large supersymmetry breaking the various contributions to the potential become of similar magnitude and 
	 couplings between the axio-dilaton, complex structure and the K\"ahler moduli are all important \cite{Gallego:2011jm,Gallego:2013qe}. Moreover, 
the applicability of the four-dimensional ${\cal N}=1$ supergravity as an effective description of the low-energy theory becomes questionable as the hierarchy between the compactification scale and the supersymmetry breaking scale decreases.  We will not be concerned about this regime in this paper.	 	
	} If the supersymmetry breaking in the complex structure and axio-dilaton sector is suitably small, however,  $F_I \bar F^I$ can provide 
	the positive energy contribution necessary to lift the LVS AdS vacua to positive vacuum energy, without causing destabilisation. 
	
	For $F_I = \epsilon  W f_I$ as in \eqref{eq:FI_LVS}, we have $F_I \bar F^I = \epsilon^2 |W^2|$ and  the condition that  at the minimum the extra contribution to the potential is of the same order as the LVS contribution requires that:
	\be
	 \epsilon = \O (\Vmin^{-1/2})\, .
	\label{eq:scaling}
	\ee  
		Such a small contribution can  be achieved by the tuning of fluxes: for continuous fluxes, such tuning is always possible; for quantised fluxes, we expect  this be possible in compactifications with many cycles and a large flux tadpole. We emphasise, however, that these new F-terms are the consequence of non-supersymmetric three-form fluxes, and not the result of an ubiquitous small backreaction from the non-SUSY solution in the K\"ahler sector, which induces even smaller F-terms of the order of $F_I \bar F^I \sim {\cal O}(|W|^2/\Vmin^2) \sim {\cal O}(\epsilon^4 |W|^2)$ in the standard Large Volume Scenario.


	To construct perturbatively metastable  vacua, we need to solve the critical point equations and ensure that the Hessian matrix has no negative eigenvalues. With non-vanishing supersymmetry breaking in the complex structure and axio-dilaton sector, the critical point equation for these moduli fields becomes,
	\be\label{eq:EOMS}
	\partial_I V = \partial_I V_{\rm LVS} + e^{\tilde K} \frac{ \left(\bF^J Z_{IJ} + F_I \bW \right)}{\V^2} \, , 
	\ee
	where we have used that $\partial_I \left(e^{\tilde K} F_J \bF^J \right) = e^{\tilde K} \left( \bF^J Z_{IJ} + F_I \bW \right)$.
	Since $\left( \bF^J Z_{IJ} + F_{I} \bW \right) \sim \Vmin^{-1/2}$ and 
	$\partial_I V_{\rm LVS}$ scales like $\Vmin^{-3}$, 			
	the second term provides the leading order contribution to the critical point equation in an expansion in the inverse volume. Thus,  up to corrections of ${\cal O}(\Vmin^{-1})$, the equation $\partial_I V=0$ implies that,\footnote{Since $Z_{SS} \equiv 0$ in general, $F_I$ cannot be exactly aligned with the axio-dilaton direction, but needs to have some non-vanishing component in the complex structure sector.   } 
	\be
	\bF^J Z_{IJ} = - \bW F_I \, .
	\label{eq:cpcs2}
	\ee
	This is precisely equation \eqref{eq:cpcs}. The implications of this equation are again very important: 
	the three-form fluxes can stabilise most of the fields in the complex structure and axio-dilaton sector with squared masses of  $ \sim 1/\Vmin^2$, but 
	a consistent inclusion of the additional supersymmetry breaking effects requires 
	one eigenvalue of $(Z \bZ)$ to be equal to $|W|^2$. So, again,  
	one real component of these fields is unstabilised at leading order. We will again denote this component by $X^{1-}$ and  we will see in section \ref{sec:Hessian} that this field is lifted at subleading order and can be stabilised by very modest amounts of tuning. 
	
	Granted a solution of  equation \eqref{eq:cpcs2} (we  give an explicit prescription for  solving it in section \ref{sec:example}), we now turn to the critical point equation for the K\"ahler moduli. The scalar potential is now given by, 
	\bea
	V&=&
	\frac{\tilde a}{\V^3} - \frac{ \sum_{s} \tilde b_s x_s e^{-x_s}  
		- \epsilon^2 \tilde f^2
	}{\V^2} + \frac{\sum_{s} \tilde c_s \sqrt{x_s} e^{-2 x_s}}{\V}
	\, ,\label{eq:Vsimp}
	\eea
	where, $\tilde a,~ \tilde b,~ \tilde c,~ \tilde f$ are complex structure and axio-dilaton dependent (but K\"ahler moduli independent) parameters:
	\bea
	\tilde a &=& \frac 34 e^{\tilde K}\,  \xi ({\rm Im}(S))^{3/2} |W_0|^2 \, , ~~~
	\tilde  b_s=4e^{\tilde K} |A_s W_0|   \, , \\
	\tilde  c_s&=& \frac{8}{3} e^{\tilde K}  \frac{a_s^{\frac32} |A_s|^2}{\eta^{3/2}_s } \,,~~~
	\epsilon^2 \tilde f^2 = e^{\tilde K} F_I \bar F^I \, .
	\eea
	Clearly, while the new $\tilde f^2 = {\rm exp}(\tilde K) |W_0|^2$ contribution affects the equation $\partial_{\V} V = 0$, it does not enter into the critical point equations for the `small' K\"ahler moduli,  $\partial_{x_s} V = 0$. Hence, we again find,
	\be
	\Vmin = 2 \frac{\tilde b_s}{\tilde c_s} e^{x_s} \sqrt{x_s} \left( \frac{x_s-1}{4x_s-1}\right)\approx \frac{\tilde b_s}{2 \tilde c_s} e^{x_s} \sqrt{x_s}
	\, ,
	\label{eq:xssol2}
	\ee
	for each $x_s$, as in equation \eqref{eq:taussol}. Equation \eqref{eq:Vsol} instead is generalised to,
	\be
	\tilde a+\frac23 \epsilon^2 \tilde f^2 \Vmin= 4 \sum_{s=1}^{N_{small}}\frac{ b_s^2}{c_s}x_s^{5/2}\frac{ (x_s-1) }{ (4 x_s-1)^2} \approx \frac 14 \sum_{s=1}^{N_{small}}\frac{b_s^2}{c_s}x_s^{3/2}\,.
	\label{eq:Vsol2}
	\ee
	
	The implicit solutions of the critical point equations (cf.~\eqref{eq:cpcs2}, \eqref{eq:xssol2} and \eqref{eq:Vsol2}) do not guarantee that the solution is a local minimum of the potential. To investigate the metastability of theses we now turn to the Hessian matrix.

	\subsection{Metastable vacua}
	\label{sec:Hessian}
	Perturbative metastability requires that all the eigenvalues of the Hessian matrix are positive. 
	A key concern is the magnitude  of the off-diagonal terms in the mass matrix, which lead to mass splittings that can cause destabilisation. To 
	estimate the importance of these effects, it often suffices to recall that to
	second order in matrix perturbation theory for the matrix ${\cal H}_{AB} = {\cal H}^{(0)}_{AB} + \delta {\cal H}_{AB}$, the perturbed eigenvalues are given by, 
	\begin{equation}
	m_A^2=m_{(0),\, A}^2+\delta {\cal H}_{AA}+\sum_{B\neq A}\frac{|\delta {\cal H}_{AB}|^2}{m_{(0),\, A}^2-m_{(0),\, B}^2}\,,
	\label{eq:2ndopt}
	\end{equation} 
	where $m_{(0),\, A}^2$ denotes the eigenvalues of the unperturbed matrix ${\cal H}^{(0)}_{AB}$. We will now show that the volume scaling of the various terms in the Hessian matrix makes ensuring perturbative stability a rather simple task.

	Most of the fields in the complex structure moduli and the axio-dilaton sector are stabilised at a high scale. The Hessian matrix to zeroth order in $\epsilon$ is again given by \eqref{eq:HessLin}, and these fields have canonically normalised squared masses,\footnote{The volume scaling of the masses in this sector is unaffected by the canonical normalisation.}
	\be
	m^2_{I\pm} = \frac{e^{\tilde K} \left(\lambda_I \pm |W_0| \right)^2}{\V^2} + {\cal O}\lp 1/\V^{5/2}\rp \, .\label{eq:masses}
	\ee
	However, since the critical point equation \eqref{eq:cpcs2} requires that $(Z\bZ)$ has one eigenvalue equal to $|W|^2$, one real degree of freedom in this sector,  $X^{1-}$, is lifted at ${\cal O}\lp1/\V^{5/2}\rp$ 
	precisely as discussed in section \ref{sec:approxnoscalereview}. The leading order squared mass is again given by,
	\be 
	m^2_{1-} = \frac{e^{\tilde K} {\rm Re}\left(U_{IJK}\bar f^I\bar f^J\bar f^K \bW \right)}{\V^2} \epsilon \, .\label{eq:lightmass}
	\ee 
	In general we expect the contribution $U_{IJK}\bar f^I\bar f^J\bar f^K$  to be order one. If this contribution is sub-leading, however, this mode will be lifted at ${\cal O}\lp1/\Vmin^3\rp$ and a more complicated analysis is required. This could occur if $D_k\left( \kappa_{ijl} e^{K_{\rm c.s.}} \right) = 0$, which holds if the moduli space is a symmetric space with a covariantly  constant Riemann  tensor. This is the case for the $STU$-model considered in \cite{Marsh:2014nla}, but for the moduli space of a more general Calabi-Yau compactification,  $D_k\left( \kappa_{ijl} e^{K_{\rm c.s.}} \right) \neq 0$.\footnote{We note  that,  contrary to an aside assertion of \cite{Denef:2004cf}, generically $D_k\left( \kappa_{ijl} e^{K_{\rm c.s.}} \right) \neq 0$ in the large complex structure limit of the compactification manifold. }
	
	The stability of the K\"ahler moduli sector can be understood from  the volume scaling of the elements of the full Hessian matrix. 
	We here, for simplicity, consider non-canonically normalised fields; canonical normalisation does not change the metastability of a critical point. 
	Cross-terms between the `small' K\"ahler moduli on the one hand and the axio-dilaton and complex structure moduli on the other arise only from $V_{\rm LVS}$ in equation \eqref{eq:Vmodified}, and hence $\partial^2_{I \tau^s} V \sim \partial_{\bar I \tau^s} V \sim 1/\V^3$. Cross-terms with the overall volume modulus scale like $\partial^2_{I \V} V \sim \partial^2_{\bar I \V} V \sim 1/\V^4$.
		
	The Hessian matrix is then schematically given by,
	\be
	{\cal H} =
	\left(
	\begin{array}{cc|cc}
		m^2_{I\pm}&  \vec{0} &\partial^2_{I \tau^s}V &\partial^2_{I \V} V\\
		\vec{0}^T &m^2_{1-} &  \partial^2_{I \tau^s}V & \partial^2_{I \V}V \\\hline
		\partial^2_{I \tau^s}V  &  \partial^2_{I \tau^s}V &\partial^2_{\tau^s \tau^s} V&\partial^2_{\tau^s \V} V\\
		\partial^2_{I \V} V & \partial^2_{I \V} V&\partial^2_{\tau^s \V} V & \partial^2_{\V \V} V \end{array}
	\right)
	=\O(\V^{-2})
	\left(
	\begin{array}{cc|cc}
		1&  \vec{0}^T & \O(\V^{-1}) & \O(\V^{-2}) \\
		\vec{0} &\O(\V^{-\frac12}) &   \O(\V^{-1}) & \O(\V^{-2})  \\\hline
		\O(\V^{-1}) &  \O(\V^{-1}) & \O(\V^{-1}) & \O(\V^{-2}) \\
		\O(\V^{-2}) & \O(\V^{-2}) & \O(\V^{-2}) &\O(\V^{-3})  \end{array}
	\right)  \,.
	\label{eq:HVscaling}
	\ee
	Using equation \eqref{eq:2ndopt}, we see that cross-terms between the complex structure and axio-dilaton sector and the K\"ahler moduli contribute only with sub-leading corrections. The `small' K\"ahler moduli squared masses are corrected at order  $\O\lp \V^{-7/2}\rp$, but the diagonal contributions are of ${\cal O}\lp \V^{-3}\rp$. The overall volume squared mass scales is corrected at  $\O\lp\V^{-11/2}\rp$, but the leading order contributions from the K\"ahler sector enter at ${\cal O}\lp \V^{-5} \rp$.
 Thus, we can consistently neglect cross-couplings between the axio-dilaton and complex structure moduli on the one hand, and the K\"ahler moduli on the other. We also note  that the volume scalings of equation \eqref{eq:HVscaling} are still satisfied for moduli dependent prefactors $A_s(S, u^i)$ of equation \eqref{eq:W},  and we expect the decoupling to apply also in this more general case.
 
Focussing on the K\"ahler moduli sector, equations \eqref{eq:2ndopt} and \eqref{eq:HVscaling} imply that the small K\"ahler moduli are stabilised at ${\cal O}(1/\V^3)$, and that cross-couplings with the overall volume modulus only lead to very small corrections. Since the potential \eqref{eq:LVSpotential} is sum-separable in the fields $\tau^s$, the only non-vanishing elements of the Hessian matrix are the diagonal values which (again for  non-canonically normalised fields) are given by,
\be
\partial^2_{\tau^s \tau^s} V|_{\rm min} =\frac{a_s^2 \tilde{b}_s^2 (x_s-1) \left(x_s \left(8 x_s^2-6 x_s+3\right)+1\right)}{\tilde{c}_s \V_{\rm min}^3 (4 x_s-1)^2 \sqrt{x_s}}\approx \frac{a_s^2 \tilde{b}_s^2 x_s^{\frac32}}{2 \tilde{c}_s \V_{\rm min}^3} \, ,
\label{eq:Vss}
\ee
with no sum on $s$. Here we have used equations \eqref{eq:xssol2} and in the last step we took $x_s \gg1$. Equation \eqref{eq:Vss} clearly implies positive squared masses for the small K\"ahler moduli, just as in the standard Large Volume Scenario. 

	We now turn to the metastability of the overall volume, and the possibility to get Minkowski or de Sitter vacua from this construction.


	\subsection{Metastability condition and de Sitter vacua}\label{sec:metstabanal}
	
	The leading order vacuum energy in the presence of $\tilde f\neq 0$ is given by, 
	\be
	V|_{\rm min} = \frac{1}{\V_{\rm min}^3}\left(\frac13 \epsilon^2 \tilde f^2\V_{\rm min}-2\sum_s\frac{\tilde b_s^2(x_s-1)x_s^{3/2}}{\tilde c_s(4x_s-1)^2}\right)\,.
	\ee
	Clearly, a non-negative vacuum energy then requires, 
	\be
	6\sum_s\frac{\tilde b_s^2(x_s-1)x_s^{3/2}}{\tilde c_s(4x_s-1)^2}\leq \epsilon^2 \tilde f^2\V_{\rm min}\,.
	\label{eq:cond1}
	\ee
	However, $\tilde f$ is also bounded from above: the lightest mode, which coincides at leading order with the overall volume modulus, is destabilised by too large flux-induced F-terms. More precisely, the second derivative of the potential with respect to the volume is given by,
	\be
	\partial^2_{\V}V|_{\rm min}=\frac{2}{\V_{\rm min}^5}\left(2\sum_s\frac{ \tilde b_s^2 x_s^{3/2} (2 x_s^2-x_s-1) }{\tilde c_s (4 x_s-1)^2}-\epsilon^2 \tilde f^2 \V_{\rm min}\right)\,.
	\ee
	The corresponding eigenvalue of the Hessian matrix gives the squared mass. For a single small blow-up modulus and non-canonically normalised fields it is given by, 
	\be\label{eq:VolumeMass2}
	m^2_{\V}=\frac{2}{\V_{\rm min}^5}\left(\frac{6 \tilde b_1^2 x_1^{5/2} \left(12 x_1^3-23 x_1^2+16 x_1-5\right)}{c_1  (4 x_1-1)^2 \left(8 x_1^3-6 x_1^2+3 x_1+1\right)}-\epsilon^2 \tilde f^2 \V_{\rm min} \rp\,.
	\ee
	 Thus, metastability requires, 
	\be
\epsilon^2 \tilde f^2 \V_{\rm min}<\frac{6 \tilde b_1^2 x_1^{5/2} \left(12 x_1^3-23 x_1^2+16 x_1-5\right)}{c_1  (4 x_1-1)^2 \left(8 x_1^3-6 x_1^2+3 x_1+1\right)}  \, .
	\label{eq:cond2}
	\ee
Defining, 
\be
h_s(x_s) = \frac{\tilde b_s^2}{\tilde c_s} \frac{(x_s-1) x_s^{3/2}}{(4x_s-1)^2}  \geq 0\, ,
\label{eq:hs}
\ee
the constraints \eqref{eq:cond1} and \eqref{eq:cond2} can be written as,
\be
1\le \frac{\epsilon^2 \tilde f^2 \V}{6 h_1(x_1)}<\frac{x_1 (5+x_1 (12 x_1-11))}{1+x_1 \left(8 x_1^2-6 x_1+3\right)}\,.
\label{eq:bound}
\ee
The right-hand-side is a monotonic function of $x_1 \geq 1$ that is equal to 1 for $x_1=1$ and asymptotes to 3/2 for $x_1 \rightarrow \infty$. Thus for any $x_1= a_1 \tau^1 >1$ there always exists a range of values for $\epsilon^2$ for which we have stable dS vacua. 

This result is easily generalised to the several blow-up moduli situation in the case we are interested in, namely when these are supported by individual non-perturbative effects.
%
%
In this case, the effects of the blow-up moduli  on the volume modulus mass simply `add up', and the generalisation of equation \eqref{eq:bound} is given by,
\be
\sum _s h_s(x_s)\le  \frac{\epsilon^2 \tilde f^2 \V}{6 }<\sum_s \left( \frac{x_s (5+x_s (12 x_s-11))}{1+x_s \left(8 x_s^2-6 x_s+3\right)}h_s(x_s)\right)\,.
\label{eq:bound2}
\ee
By the same reasoning as in the single blow-up modulus case,  there is always a permitted range of values for $\epsilon$ that gives rise to metastable de Sitter vacua. From equation \eqref{eq:bound2}, we see that this range grows with the inclusion of additional K\"ahler moduli.

	\section{An explicit example for \texorpdfstring{$\mathbb{CP}^4_{11169}$}{}}\label{sec:example}
	In this section, we give a  general method for constructing examples of the new class of metastable de Sitter vacua presented in this paper, and we illustrate this method by an explicit example. 
	
	\subsection{Method}\label{sec:method}
The class of  vacua constructed in this paper can be obtained explicitly 
in a straightforward, step-by-step procedure, at least for continuous fluxes. Key to this is the decomposition of the flux vector \eqref{eq:NPi}, which  allows us to  engineer a consistent supersymmetry breaking flux background. We here present the detailed steps of this prescription. 

{\bf Method:}
	\begin{enumerate}
		\item Pick a  point $p$ in the complex structure and axio-dilaton moduli space at which the vacuum should be realised;  evaluate the basis vectors $\{\Pv, \Pv^*, D_i \Pv, \bar D_{\ib} \Pv^*\}|_p$ at this point.
		\item Turn on fluxes along the $D_i \Pv|_p$ directions only. According to \eqref{eq:NPi}, this corresponds to freely choosing the values of $Z_{i S}|_p$. These fluxes are supersymmetric with $W_0|_p=0$, and the  flux vector is now given by, 
		\be
		\vec N =\left. -\rmi e^{\Kcs} \left(
		\frac{\bZ^i_{~\bar S}}{K_{\bar S}} D_i \Pv 
		\right)\right|_p \, .
		\label{eq:N1}
		\ee
		Given \eqref{eq:Zij}, this also specifies $Z_{ij}|_p$. In fact, since $Z_{SS} =0$  this determines the entire tensor $Z_{IJ}|_p$ in the complex structure and axio-dilaton sector. 
		

	
		\item Solve the equation,
		\be
		v^{* I} Z_{IJ}|_p = - \lambda^* v_J \, ,
		\ee
		for $v_I$ and $\lambda$, with $v_I (K^{I \bar J}|_p) v^*_{\bar J} =1$. The vector $v_J$ is then an eigenvector of $(Z\bZ)_I^{~J}|_p$ with the eigenvalue $|\lambda|^2$. 
		We now take  $W_0|_p=\lambda$ 
		 and add to the flux vector the corresponding contribution along the $\Pv^*|_p$ direction,
		\be
		\vec N = \left. -\rmi e^{\Kcs} \left(
		W_0 \Pv^* + \frac{\bZ^i_{~\bar S}}{K_{\bar S}} D_i \Pv 
		\right) \right|_p\, .
		\label{eq:N2}
		\ee
		The flux now has (2,1) and (0,3) components, and will generically induce non-vanishing F-terms for  the K\"ahler moduli, but not for the complex structure moduli or the axio-dilaton. 
		\item Fix  the flux induced contribution to the D3-tadpole, cf.~\eqref{eq:tadpole}, to any desired value by re-scaling $Z_{Si}|_p$ and
		 $W_0|_p$. We will further slightly deform this flux background, but this will only  change the tadpole by a small amount of ${\cal O}(\epsilon^2)$ that we will not be concerned with. 
		\item For the choice of branes and instantons supporting the non-perturbative effects in \eqref{eq:W} and the flux-induced superpotential folowing from \eqref{eq:GVW} and \eqref{eq:N2}, find the non-supersymmetric AdS vacuum following the Large Volume Scenario prescription reviewed in section \ref{sec:LVSreview}. Compute the magnitude of the uplift required to achieve a semi-positive definite vacuum energy. 
		\item Take $F_I = \epsilon\, W v_I|_p$ and choose $\epsilon \ll 1$ so that these F-terms can provide the right level of uplift. Add fluxes along the $\Pv|_p$ and  $\bar D_{\ib} \Pv^*|_p$ directions so that the full flux vector is given by \eqref{eq:NPi}. These fluxes are now constructed to satisfy the non-supersymmetric critical point equation \eqref{eq:cpcs}. 
		\item Minimise the full potential, dependent on both K\"ahler moduli and complex structure moduli, to find the perturbatively stable de Sitter vacuum.\footnote{Due to the coupling between the {\K} sector and the axio-dilaton and complex structure sectors, the axio-dilaton and complex structure vevs at the  de Sitter vacuum are slightly shifted from their previously chosen values at the point $p$.}
	\end{enumerate}
	
	
		\subsection{Example}
		We here illustrate our general prescription to construct  de Sitter vacua by presenting an explicit example using  the Calabi-Yau threefold
		obtained as a hypersurface in the projective space $\mathbb{CP}^4_{11169}$. This Calabi-Yau has  $h^{1,1} = 2$ and $h^{2,1} = 272$, but to  to render the model tractable, we follow the  discussion of this example in \cite{Balasubramanian:2005zx} and consistently set all but two of the complex structure moduli to zero. 
		
		\subsubsection{$\mathbb{CP}^4_{11169}$}
		The geometry of the axio-dilaton moduli space of this compactification is determined by the pre-potential which  in the `large complex structure expansion' is given by,
		\bea
		{\cal F} &=&
		-\tfrac{3}{2}(u^1)^3-\tfrac{3}{2}(u^1)^2u^2-\tfrac{1}{2}u^1(u^2)^2+\tfrac{9}{4}(u^1)^2+\tfrac{3}{2}u^1u^2 + \tfrac{17}{4}u^1+\tfrac{3}{2}u^2-\rmi\zeta(3)\tfrac{135}{4\pi^3} \, .\quad
		\eea
		In addition, the prepotential receives instanton corrections that are exponentially suppressed for sufficiently large ${\rm Im}(u^i)$, and  will not be important for our example. The {\K} potential for the complex structure moduli is then explicitly given by,
		\bea
		\Kcs&=&-\log \ls\frac{\rmi}{2} (u^1-\bar u^1)\! \left(3 (u^1-\bar u^1) (u^2-\bar u^2)+3 (u^1 - \bar u^1)^2 + (u^2-\bar u^2)^2 \right)\! + \frac{135 \zeta (3)}{\pi ^3}\rs
	\, .	\nonumber
	\eea
The compactification volume is expressed in terms of four-cycle moduli as, 
		\be
		\V= \frac{1}{9 \sqrt{2}}\lp \left(-\rmi\frac{T^{\rm big}-\bar{T}^{\rm big}}{2}\right)^{3/2}-\left(-\rmi \frac{T^s-\bar T^s}{2} \right)^{3/2}\rp\, ,
		\ee
		and the K\"ahler potential is given by equation \eqref{eq:KK} with  $\xi =- \frac{\chi(CY_3) \zeta(3)}{2 (2\pi)^3} \approx 1.31$.

		The  flux-induced superpotential is  given by,
		\bea
		W_0&=&\frac{1}{4} (h_0 S-f_0) \left(6 (u^1)^3+6 (u^1)^2 u^2+u^1 \left(2 (u^2)^2+17\right)+6 \left(u^2-\frac{45 \rmi \zeta (3)}{\pi ^3}\right)\right)\cr
		&&+\frac{1}{4}(f_1-h_1 S) \left(18 (u^1)^2+6 u^1 (2 u^2-3)+2 (u^2)^2-6 u^2-17\right) \cr
		&&+\frac{1}{2} (f_2-h_2 S) \left(3 (u^1)^2+u^1 (2 u^2-3)-3\right)+ f_3-h_3 S+(f_4-h_4 S)u^1 \cr
		&&+ (f_5-h_5 S)u^2
		\, ,
		\eea
		and  non-perturbative effects are assumed to be supported by a stack of ten D7-branes wrapping the blow-up cycle; we take, 
		\be
		a_s = \frac{2 \pi}{10}\, , \qquad  A_s = 1\,.\label{eq:aA}
		\ee

\subsubsection{Step-by-step procedure}
We here give a very detailed description of the construction of a single de Sitter vacuum in $\mathbb{CP}^4_{11169}$. Throughout this example, we work with continuous fluxes and present our results with three significant digits, however, we keep working at each step with numbers with higher precision.

\begin{enumerate}
\item We choose to construct a vacuum at $p = \{S=u^1=u^2 = 2\rmi\}$, i.e.~we set the axion vevs in this sector to zero. The relevant period vectors are, 
\bea
\Pv &\approx& \left(
\begin{array}{c}
 1 \\
 2 \rmi \\
 2 \rmi \\
  -19.1 \rmi \\
 36.3 +12 \rmi \\
 11.5 +3 \rmi \\
\end{array}
\right)
 , \,\,\,
 D_1 \Pv \approx 
\left(
\begin{array}{c}
 0.558 \rmi \\
 -0.117 \\
 -1.12 \\
 -17.1 \\
 -2.20-3.76 \rmi \\
 -0.175-1.58 \rmi \\
\end{array}
\right), \,\, \,
 D_2 \Pv \approx
\left(
\begin{array}{c}
 0.174 \rmi \\
 -0.349 \\
 0.651 \\
 -5.16 \\
 -0.594-1.67 \rmi \\
 -0.523+0.00669 \rmi \\
\end{array}
\right). \nonumber 
\eea

\item We choose to turn on fluxes so that $Z_{S1}|_p= 2.4$ and $Z_{S2}|_p = -5.43$. The full tensor $Z_{IJ}|_p$ is then given by,
\be
Z_{IJ}|_p \approx 
\left(
\begin{array}{ccc}
 0 & 2.4 & -5.43 \\
 2.4 & -7.24 & -2.41 \\
 -5.43 & -2.41 & 5.57 \\
\end{array}
\right) \, .
\ee
The flux vector is now given by equation \eqref{eq:N1}. The  RR and NSNS fluxes expressed in the original basis are given by, 
\be
\vec{f} \approx
\left(
\begin{array}{c}
 0 \\
 6.03 \\
 -18.7 \\
 3.65 \\
 -0.9 \\
 9.04 \\
\end{array}
\right) \, ,
\qquad
\vec{h} \approx
\left(
\begin{array}{c}
 0.113 \\
 0 \\
 0 \\
 0 \\
 -5.55 \\
 4.46 \\
\end{array}
\right) \, .
\ee
By construction, these fluxes give $W|_p= F_I|_p =0$.
\item The equation $v^{* I} Z_{IJ}|_p = - \lambda^{*} v_J$ has several solutions, and $\lambda$ can be chosen as the positive or negative square-root for either of the three eigenvalues of $(Z \bZ)_I^{~J}|_p$. We take $\lambda \equiv \lambda_1 \approx -104$, corresponding to the middle eigenvalue, the others being $\lambda_2 \approx -53.6$ and $\lambda_3 \approx-1230$. The vector $v_I$ is then given by,
\be\label{eq:explEiVec}
v_I \approx 
\left(
\begin{array}{c}
 -0.196 \rmi \\
 +0.189 \rmi \\
 +0.0365 \rmi \\
\end{array}
\right) \, ,
\ee
which is unit normalised with respect to the field space metric. We identify our chosen value of $\lambda$ with the desired vacuum expectation value of the flux superpotential, and turn on (0,3) fluxes according to \eqref{eq:N2}. The RR and NSNS fluxes are now given by,
\be
\vec f \approx
\left(
\begin{array}{c}
 0 \\
 6.93 \\
 -17.8 \\
 -5.03 \\
 4.55 \\
 10.4 \\
\end{array}
\right) \, , \qquad 
\vec  h \approx 
\left(
\begin{array}{c}
 -0.114 \\
 0 \\
 0 \\
 0 \\
 -13.8 \\
 1.85 \\
\end{array}
\right) \, .
\ee
\item In this example, the flux induced tadpole becomes,
\be 
(2\pi)^4 (\alpha')^2 Q_{\rm flux} = \vec{h}^T \Sigma \vec{f}\approx 129\,.
\ee
This is (by our ostensibly prescient choice of $Z_{Si}$) already consistent with the tadpole constraint from a known embedding of $\mathbb{CP}^4_{11169}$ into F-theory \cite{Denef:2004dm}. However, due to the linearity of the problem, other tadpole constraints corresponding to other 7-brane configurations can be satisfied by simply rescaling $W_0|_p$ and $Z_{Si}|_p$. 
\item To estimate the magnitude of the required uplift, we minimise the LVS scalar potential \eqref{eq:LVSpotential} for the K\"ahler moduli sector alone, given $W_0|_p=-104$,\footnote{In the literature (for example \cite{Kachru:2003aw}),  a commonly used definition of $W_0$ used in studies of   K\"ahler moduli  stabilisation includes a normalisation factor from 
the complex structure and axio-dilaton K\"ahler potential. In our example, 
 $\left(e^{\tilde K/2} W_0 \right)\big|_p= -3.44$.} and find  a non-supersymmetric AdS solution.\footnote{By construction, this solution has a leading-order flat direction in the axio-dilaton and complex structure sector.} In this example,  the volume of the AdS vacuum is given by,
$
\V|_p \approx 6360
$ and the leading order `cosmological constant' is given by,
\be\label{eq:AdSCC}
V|_p \approx -1.15\times 10^{-11} \, .
\ee 
In point 6 below, the inclusion of non-vanishing F-terms for the complex structure moduli and the axio-dilaton will shift the vev of the volume.  
%
Nevertheless, the  volume 
of the AdS vacuum indicates the rough size of the desired uplift: $\epsilon \sim \V^{-1/2}|_p\approx 1/80$.
\item We now turn on fluxes that are, at $p$, not supersymmetric in the complex structure and axio-dilaton sector. To satisfy the leading order critical point equation, cf.~equation \eqref{eq:cpcs}, these have to be proportional to $v_I$. 
We choose $\epsilon^2 = F_I \bar F^I/|W|^2 = (1/120)^2$, and find the slightly corrected fluxes,
\be
\vec f \approx
\left(
\begin{array}{c}
 0 \\
 6.94 \\
 -17.7 \\
 -4.98 \\
 4.60 \\
 10.4 \\
\end{array}
\right)\, , \qquad
\vec h \approx
\left(
\begin{array}{c}
 -0.114 \\
 0 \\
 0 \\
 0 \\
 -13.7 \\
 1.88\\
\end{array}
\right) \, .
\label{eq:fluxFinal}
\ee

\item Finally, we directly minimise the full scalar potential given the fluxes \eqref{eq:fluxFinal}. We consistently find a critical point near, but not exactly at $p$, with relative deviations of ${\cal O}(\V^{-1/2})$ due to the mixing with the K\"ahler moduli. The moduli vevs at the critical point are given by,
\bea
	S &\approx& 1.96 \,\rmi \, , \quad  u^1 \approx 2.03 \, \rmi \, , \quad u^2 \approx 1.97\, \rmi \, , \\
\tau^s &\approx& 9.64 \, ,  \quad \V \approx 11 300\, .
\eea
The `small' K\"ahler modulus axion is stabilised at zero (since $W_0 < 0$ and $A_s >1$), and the axion of the `large' cycle remains unfixed. The vacuum expectation value of the potential at the critical point is positive,
\be
V \approx 9.03 \times 10^{-14} \, .
\ee
The eigenvalues for the Hessian matrix for canonically normalised fields are given in tables \ref{tab:SUsectorcanmass} and \ref{tab:masscanKahler}. The spectrum is positive semi-definite, with only the  axion of the large {\K} modulus still massless. 
		
		\begin{table}[hbt]
			\begin{center}
				\begin{tabular}{|c|c|c|c|c|c|}\hline 
					$m_{3+}^2$ & $m_{3-}^2$ & $m_{1+}^2$ & $m_{2+}^2$ & $m_{2-}^2$ & $m_{1-}^2$  \\ \hline
					$1.61 \times 10^{-5}$ &  $1.15 \times 10^{-5}$ & $3.56 \times 10^{-7}$ & $2.12 \times 10^{-7}$ & $2.20 \times 10^{-8}$ & $2.23 \times 10^{-9}$ \\ \hline
				\end{tabular}
			\end{center}
			\caption{Canonically normalised squared masses  for the axio-dilaton and complex structure moduli in natural units. We labeled the masses squared as in section \ref{sec:review1stpaper}.}\label{tab:SUsectorcanmass}
		\end{table}

\begin{table}[htb]
	\begin{center}
		\begin{tabular}{|c|c|c|c|}\hline
			$m_{\Re(T^s)}^2$& $m_{\Im(T^s)}^2$ & $m_{\Im(T^{\rm big})}^2$ & $m_{\Re(T^{\rm big})}^2$ \\\hline
			$1.18\times 10^{-5}$ & $1.10 \times 10^{-5}$ &  $7.06 \times 10^{-12}$ & $0$\\\hline
		\end{tabular}
	\end{center}
	\caption{Canonically normalised squared masses for the {\K} moduli. As in standard LVS, the masses of $T^s$ are of the same order as the masses in the axio-dilaton and complex structure sector. As previously discussed in subsection \ref{sec:metstabanal}, there is a leading order mass mixing between the overall volume and the small cycle, so that the eigenvalue denoted by $m_{\Im(T^{\rm big})}^2$ corresponds to an eigenvector with leading entries along the $\Im(T^{\rm big})$ and $\Im(T^s)$ directions. }\label{tab:masscanKahler}
\end{table}

\end{enumerate}

We close this section by discussing the consistency of our analytical discussion in section \ref{sec:dS} with the spectra of tables \ref{tab:SUsectorcanmass} and \ref{tab:masscanKahler}.

The canonically normalised masses for the axio-dilaton and complex structure moduli are well described by the analytical expressions reported in \cite{Marsh:2014nla} and reviewed in sections \ref{sec:review1stpaper} and \ref{sec:dS}, in particular equations \eqref{eq:masses} and \eqref{eq:lightmass} reproduce the masses squared in table \ref{tab:SUsectorcanmass} at the percent level. 

Similarly, we can compare the squared  masses for the small {\K} modulus with the expression given in \eqref{eq:Vss}. For a single small {\K} modulus and without the assumption $x_s \gg 1$, we find the canonically normalised mass squared
\be
\frac12 K^{T^s \bar{T}^s} \partial^2_{\tau^s \tau^s} V \approx 1.10 \times 10^{-5} \,,
\ee
which is in perfect agreement with the result above. 

As a final check of our solution, we can calculate the canonically normalised mass for the overall volume from the simplified scalar potential given in equation \eqref{eq:Vsimp}. This requires us to determine $\epsilon^2 \tilde f^2 = e^{\tilde K} F_I \bar F^I$. We find for our example,
 \be
  e^{\tilde K} F_I \bar F^I= 0.00109\lp 0.793+\frac{828}{\V}+\frac{314000}{\V^2}\rp \, ,
  \ee
   where the terms that are subleading in powers of the  volume arise from the inverse {\K} metric in $F_I \bar F^I = F_I K^{I \bar J} \bar F_{\bar J}$. 
   However, the potential of equation \eqref{eq:Vsimp} is only valid to leading order in the volume, and keeping only the leading order contribution to $  e^{\tilde K} F_I \bar F^I$  we have  $\epsilon^2 \tilde f^2 = e^{\tilde K} F_I \bar F^I =0.000866$. We then find for the volume modulus the canonically normalised squared mass\footnote{As discussed in subsection \ref{sec:metstabanal}, there is a non-negligible mass mixing with the small cycle so we have to diagonalise the Hessian. Note, that the eigenvalue given in \eqref{eq:VolumeMass2} corresponds to non-canonically normalised fields so it does not give the value we quote here.} $m^2_{{\rm Im}(T^{\rm big})} \approx 1.50 \times 10^{-12} $. This value is still roughly consistent with our full result above,  but 
   shows the largest deviation from the true value than our analytical estimates of the other moduli masses. 
   This is explained by the only moderately large  volume in our explicit example, and the fact that the volume modulus is the lightest modulus and hence the most sensitive one to small corrections that arise due to the large volume approximation in section \ref{sec:dS}.

Comparing with the metastability analysis in subsection \ref{sec:metstabanal}, we find from equation \eqref{eq:bound} for the final values the range,
\be
9.38 < \epsilon^2 \tilde{f}^2 \V <  13.6\, ,
\ee
where for our solution we have $\epsilon^2 \tilde{f}^2 \V\approx 10.7$.
		
As discussed above, we did not try to get the fluxes to be simultaneously integer quantised 
and consistent with a given, moderately large flux tadpole. We expect that the flux tuning required to 
make $\epsilon \ll 1$
can be realised using quantised fluxes in compactifications with more moduli (or substantially larger flux tadpoles) than the example we have considered. 
	
	\section{\label{sec:visiblesector} Visible sector soft terms}
	In this paper we have focussed on the generation of metastable de Sitter vacua with stabilised moduli, but compactifications of string theory hoping to describe the real world  also need to include matter and gauge fields compatible with the Standard Model of particle physics. Such a `visible sector' may for example be generated by intersecting branes or by localising branes at singularities in the compactification geometry. 
	Our mechanism for generating metastable de Sitter vacua  has important consequences for the properties of the 
	%
	soft supersymmetry breaking terms in the visible sector, as we briefly outline in this section. 
		
	For concreteness, we focus on the well-studied case of visible sectors generated by branes at singularities, cf.~\cite{Blumenhagen:2009gk, Krippendorf:2010hj, Dolan:2011qu, Cicoli:2012vw, Cicoli:2013cha, Cicoli:2013mpa} for earlier work on such models in the Large Volume Scenario. We denote  the K\"ahler modulus that resolves the cycle by $T^{\rm SM}$, with the four-cycle volume controlled by ${\rm Im}(T^{\rm SM}) = \tau^{\rm SM}$. Due to the issues of large-scale breaking of the Standard Model symmetries  raised in \cite{Blumenhagen:2007sm}, the new modulus $T^{\rm SM}$ is taken to be distinct from the moduli $T_s$ that are stabilised by non-perturbative effects. The modulus $T^{\rm SM}$ is instead assumed to be stabilised supersymmetrically using D-terms, giving $F_{T^{\rm SM}} = 0$. In this case, the soft supersymmetry breaking parameters determining the low-energy phenomenology of the model are generated 	by couplings to the supersymmetry breaking bulk moduli, $T^{\rm big}$, $u^i$ and $S$, which can be computed using the general formulae of \cite{Kaplunovsky:1993rd, Brignole:1993dj, Soni:1983rm}. 
	
	We begin by computing the induced gaugino masses in our scenario. For branes localised at a collapsed singularity,  the holomorphic gauge kinetic function is given by,
	\be
	\mathfrak{f} = c_1\, S + c_2 \, T^{\rm SM} \, .
	\ee
	The gaugino masses are then given by, 
	\be
	M_{1/2}=\frac{1}{2{\rm Im}({\mathfrak{f}})}e^{K/2} \bF^A\partial_A \mathfrak{f}
	=
	\frac{1}{2{\rm Im}({\mathfrak{f}})}e^{K/2} \bF^S\partial_S \mathfrak{f}
	\sim \frac{m_{3/2}}{\Vmin^{1/2}}
	\, .
	\ee	
	While suppressed with respect to $m_{3/2}$,  these fields  are heavier by a factor of $\sqrt{\Vmin}$ than usually assumed in the Large Volume Scenario. 

For chiral visible sector fields $C^a$ with a diagonal matter metric,
\be
K = \sum_{\alpha} \tilde K_{\alpha} C^{\alpha} \bar C^{\bar \alpha} + \ldots \, ,
\ee
the 
soft masses for the non-canonically normalised fields in an approximately Minkowski vacuum are given by \cite{Brignole:1993dj},
\be
	m_\alpha^2= m_{3/2}^2  -e^K \bar F^A  F^{\bar B}\partial^2_{A \bar B} {\rm log}(\tilde K_\alpha) \,.
	\label{eq:softmass}
\ee
Thus, determining the soft masses requires knowing the moduli dependence of the matter K\"ahler metric. 

Arguments based on holomorphicity and locality suggest that for visible sectors realised at branes at singularities \cite{Conlon:2006tj}, 
\be
	\tilde K_\alpha=h_\alpha(X^I,\, \bar X^{\bar I})\, e^{K/3}  \kappa_{\alpha} \, ,
	\label{eq:sortofseq}
	\ee
for $X^I = (S,u^i)$. Here $\kappa_{\alpha}$ is assumed  to not  depend on bulk moduli. For $h_{\alpha}$ a moduli-independent constant, this K\"ahler metric is of the `sort-of-sequestered' form \cite{Berg:2010ha}, implying vanishing soft masses, just as in the scenario of sequestered supersymmetry breaking of reference \cite{Randall:1998uk}. However, sequestering in string compactifications is expected to be at best approximate,   
and non-vanishing soft terms can be induced by e.g.~$\alpha'$ corrections \cite{Blumenhagen:2009gk,Aparicio:2014wxa}, superpotential de-sequestering \cite{Berg:2010ha, Berg:2012aq}, or, in our case, a non-constant function $h_{\alpha}(X^I,\, \bar X^{\bar I})$. These de-sequestering effects can dominate over  anomaly mediated contributions to the soft masses. Explicit computations of the matter metric in certain toroidal orbifolds indicate that, at least in these cases, $h_{\alpha}$ is a non-constant function \cite{Ibanez:1998rf, Lust:2004cx, Lust:2004fi}.

For $h_{\alpha}$ non-constant, the soft terms are generated as, schematically, 
\be
m^2_{\alpha} \sim \frac{m_{3/2}^2}{\Vmin} 
\left( 
\frac{\partial \bar \partial h_{\alpha}}{h_{\alpha}} 
- \frac{\partial h_{\alpha} \bar \partial h_{\alpha}}{h_{\alpha}^2} 
\right)
 \sim M_{1/2}^2 \, .
\label{eq:msoft}
\ee
In this case, we also expect that,\footnote{
For $h_{\alpha}$ constant and 
vanishing  superpotential cross-couplings between the supersymmetry breaking moduli and the matter fields, 
equation  \eqref{eq:sortofseq} also implies the 
 vanishing of the leading order contribution to the soft  A-terms and, in the absence of a $\mu$-term, also a vanishing soft $B\mu$ term.  
}
 \be
 (A_{\alpha \beta \gamma})^2 \sim B\mu \sim \frac{m_{3/2}^2}{\Vmin} \sim M_{1/2}^2 \, .
 \label{eq:Asoft}
 \ee 	

This scenario has far-reaching phenomenological implications.\footnote{We are very grateful to Michele Cicoli for discussions on this point. }
By virtue of coupling with gravitational strength interactions to all sectors of the theory, moduli tend to be displaced from their vacuum expectation values during inflation, and subsequently come to oscillate around the vacuum, red-shifting like matter before  eventually decaying. This way, long-lived moduli can come to dominate the energy density of the universe, but -- if too long-lived -- may fail to reheat the Standard Model at temperatures sufficient for Big Bang Nucleosynthesis. This is the `cosmological moduli problem' \cite{Coughlan:1983ci,Banks:1993en,deCarlos:1993wie}. For moduli with generic, gravitational strength interactions, the corresponding bound on the decay rate implies that $m \gtrsim 50\, {\rm TeV}$. The cosmological moduli problem applies in particular to the overall volume modulus, so that,
\be
m_{\V} \sim \frac{m_{3/2}}{\sqrt{\Vmin}} \gtrsim 50\; {\rm TeV} \, .
\ee
However, according to equations \eqref{eq:msoft} and \eqref{eq:Asoft}, the soft terms are all at the same scale as the volume modulus, $m_{\V} \sim m_{\rm soft} \sim A \sim \sqrt{B\mu}$, implying that,
\be
m_{\rm soft} \gtrsim 50\; {\rm TeV} \, .
\label{eq:msoftbound}
\ee
This lower bound on the soft masses has 
obvious implications  for laboratory searches for supersymmetry,
but also constrains  the particle nature of dark matter. 
The lightest superparticle (LSP) in supersymmetric extensions of the Standard Model is the prime WIMP candidate, but in our scenario equation \eqref{eq:msoftbound} implies that it would hardly lie in the WIMP window. 
Hence, dark matter  cannot be the standard WIMP, but may well be realised by e.g.~axions.

	\section{\label{sec:conclusions} Conclusions}
	In this paper we have proposed an extension of the standard Large Volume Scenario in which we
	relaxed the assumption on exactly vanishing F-terms 
	%
	  in the axio-dilaton and complex structure sector. 
	The additional amount of supersymmetry breaking in this sector has to be small, $F^2 \sim \O(1/\V)$, in order to not destabilise the {\K} sector. 
	We showed that the non-supersymmetric critical point equation for the axio-dilaton and the complex structure moduli
	has direct implications for the moduli spectrum of the compactification, and forces one real field in this sector to be lighter than in the standard Large Volume Scenario. Just as in \cite{Marsh:2014nla} however, this field can be stabilised by a moderate tuning of a phase. 
	This ensures that  the decoupling of the axio-dilaton and complex structure sector from the {\K} sector 
	is still essentially the same as in the standard Large Volume Scenario. 
	
	The inclusion of additional sources of spontaneous supersymmetry breaking  leads to the following schematic potential for the {\K} moduli: 
	\be
	V \sim \frac{F^2}{\V^2} +\frac{\tilde a}{\V^3}
	-\frac{\tilde b \t e^{-a\t}}{\V^2}
	+ \frac{\tilde c \sqrt{\t}e^{-2 a \tau}}{\V} \, .
	\ee
	We find that the new F-term breaking can lead, for a finite range of $F^2$, to metastable de Sitter vacua that do not require any additional uplift (like for example an anti-D3-brane). Thus our construction constitutes a new class of de Sitter vacua in string theory.
	
	In section \ref{sec:example}, we presented a general method for explicitly constructing examples of this class of vacua, and we illustrated this method for the particular case of compactifications on the Calabi-Yau constructed as a hypersurface in  $\mathbb{CP}^4_{11169}$. This method relies on the continuous flux approximation, but we expect that examples with quantised fluxes can be constructed in compactifications with a larger number of three-cycles and a large flux tadpole. 
	It would be very interesting to explicitly construct such examples. 
		
	
	We also discussed potential implication for supersymmetry breaking 
	soft terms for visible sectors realised 
	through branes 
	at local singularities \cite{Cicoli:2013cha, Aparicio:2014wxa}. 
	Our solutions invoke 
	larger supersymmetry breaking contributions from the axio-dilaton and complex structure fields compared to the usual LVS construction.
	This
	  leads to larger 
	  gaugino masses than usually assumed in the Large Volume Scenario,
	  \be
	  M_{1/2} \sim \frac{m_{3/2}}{\sqrt{\Vmin}} \, , 
	  \ee
	  and soft terms of the same order,
	  \be
	  m^2_{\rm soft} \sim M^2_{1/2} \sim A^2 \sim B\mu \, . 
	  \ee 
	 Since also $m_{\cal \V} \sim m_{\rm soft}$, the resolution of the cosmological moduli problem suggest  that these soft terms are  $\gtrsim 50\, {\rm TeV}$, which predicts  null results at present laboratory searches for supersymmetry, and a non-WIMP origin of dark matter.  
	It would be interesting to study the phenomenology and cosmology of this scenario  in more detail.

	Our present analysis applies to Calabi-Yau manifolds of strong Swiss-cheese type with an arbitrary number of small cycles. It would be interesting to check whether it also extends to more general compactification topologies. 
	
	\acknowledgments
	We would like to thank  M.~Dias, J.~Frazer, R.~Kallosh, L.~McAllister, F.~Quevedo, K.~Sousa  and  A.~Westphal for illuminating discussions. We are grateful to C.~Brodie and M.~Cicoli for comments on the draft and for discussions.  
	DG thanks the Institute for Theoretical Physics, TU Wien, for hospitality and The Abdus Salam International Centre for Theoretical Physics for hospitality and support.
	BV and TW thank the Galileo Galilei Institute for Theoretical Physics for the hospitality and the INFN for partial support. 
	DM and TW are grateful for the hospitality of the Aspen Center for Physics, which is supported by National Science Foundation grant PHY-1607611. BV, DM and TW are grateful to the Lorentz center for hospitality during the completion of this work.   
	DG was partially supported by UPTC-DIN grant no.\ SGI-2145.
	DM is supported by a Stephen Hawking Advanced Fellowship at the Centre for Theoretical Cosmology at the University of Cambridge.
	 BV was supported during this work by: the European Commission through the Marie Curie Intra-European fellowship 328652--QM--sing and Starting Grant  of the European Research Council (ERC-2011-SrG 279617 TOI), in part by the Interuniversity Attraction Poles Programme initiated by the Belgian Science Policy (P7/37) and in part by the European Research Council grant no.\ ERC-2013-CoG 616732 HoloQosmos, and the KU Leuven C1 grant Horizons in High-Energy Physics.
	TW is supported by the Austrian Science Fund (FWF): P 28552.

	
	\bibliographystyle{JHEP}
	\bibliography{refs}

\providecommand{\href}[2]{#2}\begingroup\raggedright\begin{thebibliography}{10}

\bibitem{Grana:2005jc}
M.~Grana, \emph{{Flux compactifications in string theory: A Comprehensive
  review}}, \href{http://dx.doi.org/10.1016/j.physrep.2005.10.008}{\emph{Phys.
  Rept.} {\bfseries 423} (2006) 91--158},
  [\href{https://arxiv.org/abs/hep-th/0509003}{{\ttfamily hep-th/0509003}}].

\bibitem{Douglas:2006es}
M.~R. Douglas and S.~Kachru, \emph{{Flux compactification}},
  \href{http://dx.doi.org/10.1103/RevModPhys.79.733}{\emph{Rev. Mod. Phys.}
  {\bfseries 79} (2007) 733--796},
  [\href{https://arxiv.org/abs/hep-th/0610102}{{\ttfamily hep-th/0610102}}].

\bibitem{Denef:2008wq}
F.~Denef, \emph{{Les Houches Lectures on Constructing String Vacua}},  in
  \emph{{String theory and the real world: From particle physics to
  astrophysics. Proceedings, Summer School in Theoretical Physics, 87th
  Session, Les Houches, France, July 2-27, 2007}}, pp.~483--610, 2008.
\newblock \href{https://arxiv.org/abs/0803.1194}{{\ttfamily 0803.1194}}.

\bibitem{Douglas:2003um}
M.~R. Douglas, \emph{{The Statistics of string / M theory vacua}},
  \href{http://dx.doi.org/10.1088/1126-6708/2003/05/046}{\emph{JHEP} {\bfseries
  05} (2003) 046}, [\href{https://arxiv.org/abs/hep-th/0303194}{{\ttfamily
  hep-th/0303194}}].

\bibitem{Ashok:2003gk}
S.~Ashok and M.~R. Douglas, \emph{{Counting flux vacua}},
  \href{http://dx.doi.org/10.1088/1126-6708/2004/01/060}{\emph{JHEP} {\bfseries
  01} (2004) 060}, [\href{https://arxiv.org/abs/hep-th/0307049}{{\ttfamily
  hep-th/0307049}}].

\bibitem{Denef:2004ze}
F.~Denef and M.~R. Douglas, \emph{{Distributions of flux vacua}},
  \href{http://dx.doi.org/10.1088/1126-6708/2004/05/072}{\emph{JHEP} {\bfseries
  05} (2004) 072}, [\href{https://arxiv.org/abs/hep-th/0404116}{{\ttfamily
  hep-th/0404116}}].

\bibitem{Denef:2004cf}
F.~Denef and M.~R. Douglas, \emph{{Distributions of nonsupersymmetric flux
  vacua}}, \href{http://dx.doi.org/10.1088/1126-6708/2005/03/061}{\emph{JHEP}
  {\bfseries 03} (2005) 061},
  [\href{https://arxiv.org/abs/hep-th/0411183}{{\ttfamily hep-th/0411183}}].

\bibitem{Kreuzer:2000xy}
M.~Kreuzer and H.~Skarke, \emph{{Complete classification of reflexive polyhedra
  in four-dimensions}}, {\emph{Adv. Theor. Math. Phys.} {\bfseries 4} (2002)
  1209--1230}, [\href{https://arxiv.org/abs/hep-th/0002240}{{\ttfamily
  hep-th/0002240}}].

\bibitem{Lynker:1998pb}
M.~Lynker, R.~Schimmrigk and A.~Wisskirchen, \emph{{Landau-Ginzburg vacua of
  string, M theory and F theory at c = 12}},
  \href{http://dx.doi.org/10.1016/S0550-3213(99)00204-7}{\emph{Nucl. Phys.}
  {\bfseries B550} (1999) 123--150},
  [\href{https://arxiv.org/abs/hep-th/9812195}{{\ttfamily hep-th/9812195}}].

\bibitem{Riess:1998cb}
{\scshape Supernova Search Team} collaboration, A.~G. Riess et~al.,
  \emph{{Observational evidence from supernovae for an accelerating universe
  and a cosmological constant}},
  \href{http://dx.doi.org/10.1086/300499}{\emph{Astron.J.} {\bfseries 116}
  (1998) 1009--1038}, [\href{https://arxiv.org/abs/astro-ph/9805201}{{\ttfamily
  astro-ph/9805201}}].

\bibitem{Perlmutter:1998np}
{\scshape Supernova Cosmology Project} collaboration, S.~Perlmutter et~al.,
  \emph{{Measurements of Omega and Lambda from 42 high redshift supernovae}},
  \href{http://dx.doi.org/10.1086/307221}{\emph{Astrophys.J.} {\bfseries 517}
  (1999) 565--586}, [\href{https://arxiv.org/abs/astro-ph/9812133}{{\ttfamily
  astro-ph/9812133}}].

\bibitem{Aazami:2005jf}
A.~Aazami and R.~Easther, \emph{{Cosmology from random multifield potentials}},
  \href{http://dx.doi.org/10.1088/1475-7516/2006/03/013}{\emph{JCAP} {\bfseries
  0603} (2006) 013}, [\href{https://arxiv.org/abs/hep-th/0512050}{{\ttfamily
  hep-th/0512050}}].

\bibitem{Marsh:2011aa}
D.~Marsh, L.~McAllister and T.~Wrase, \emph{{The Wasteland of Random
  Supergravities}},
  \href{http://dx.doi.org/10.1007/JHEP03(2012)102}{\emph{JHEP} {\bfseries 1203}
  (2012) 102}, [\href{https://arxiv.org/abs/1112.3034}{{\ttfamily 1112.3034}}].

\bibitem{Bachlechner:2014rqa}
T.~C. Bachlechner, \emph{{On Gaussian Random Supergravity}},
  \href{http://dx.doi.org/10.1007/JHEP04(2014)054}{\emph{JHEP} {\bfseries 04}
  (2014) 054}, [\href{https://arxiv.org/abs/1401.6187}{{\ttfamily 1401.6187}}].

\bibitem{Bray:2007tf}
A.~J. Bray and D.~S. Dean, \emph{{Statistics of critical points of Gaussian
  fields on large-dimensional spaces}},
  \href{http://dx.doi.org/10.1103/PhysRevLett.98.150201}{\emph{Phys. Rev.
  Lett.} {\bfseries 98} (2007) 150201}.

\bibitem{Easther:2016ire}
R.~Easther, A.~H. Guth and A.~Masoumi, \emph{{Counting Vacua in Random
  Landscapes}},  \href{https://arxiv.org/abs/1612.05224}{{\ttfamily
  1612.05224}}.

\bibitem{Ferrara:2014kva}
S.~Ferrara, R.~Kallosh and A.~Linde, \emph{{Cosmology with Nilpotent
  Superfields}}, \href{http://dx.doi.org/10.1007/JHEP10(2014)143}{\emph{JHEP}
  {\bfseries 1410} (2014) 143},
  [\href{https://arxiv.org/abs/1408.4096}{{\ttfamily 1408.4096}}].

\bibitem{Kallosh:2014wsa}
R.~Kallosh and T.~Wrase, \emph{{Emergence of Spontaneously Broken Supersymmetry
  on an Anti-D3-Brane in KKLT dS Vacua}},
  \href{http://dx.doi.org/10.1007/JHEP12(2014)117}{\emph{JHEP} {\bfseries 12}
  (2014) 117}, [\href{https://arxiv.org/abs/1411.1121}{{\ttfamily 1411.1121}}].

\bibitem{Bergshoeff:2015jxa}
E.~A. Bergshoeff, K.~Dasgupta, R.~Kallosh, A.~Van~Proeyen and T.~Wrase,
  \emph{{$ \overline{\mathrm{D}3} $ and dS}},
  \href{http://dx.doi.org/10.1007/JHEP05(2015)058}{\emph{JHEP} {\bfseries 05}
  (2015) 058}, [\href{https://arxiv.org/abs/1502.07627}{{\ttfamily
  1502.07627}}].

\bibitem{Kallosh:2015nia}
R.~Kallosh, F.~Quevedo and A.~M. Uranga, \emph{{String Theory Realizations of
  the Nilpotent Goldstino}},
  \href{http://dx.doi.org/10.1007/JHEP12(2015)039}{\emph{JHEP} {\bfseries 12}
  (2015) 039}, [\href{https://arxiv.org/abs/1507.07556}{{\ttfamily
  1507.07556}}].

\bibitem{Aparicio:2015psl}
L.~Aparicio, F.~Quevedo and R.~Valandro, \emph{{Moduli Stabilisation with
  Nilpotent Goldstino: Vacuum Structure and SUSY Breaking}},
  \href{http://dx.doi.org/10.1007/JHEP03(2016)036}{\emph{JHEP} {\bfseries 03}
  (2016) 036}, [\href{https://arxiv.org/abs/1511.08105}{{\ttfamily
  1511.08105}}].

\bibitem{Garcia-Etxebarria:2015lif}
I.~Garc\'ia-Etxebarria, F.~Quevedo and R.~Valandro, \emph{{Global String
  Embeddings for the Nilpotent Goldstino}},
  \href{http://dx.doi.org/10.1007/JHEP02(2016)148}{\emph{JHEP} {\bfseries 02}
  (2016) 148}, [\href{https://arxiv.org/abs/1512.06926}{{\ttfamily
  1512.06926}}].

\bibitem{Dasgupta:2016prs}
K.~Dasgupta, M.~Emelin and E.~McDonough, \emph{{Fermions on the antibrane:
  Higher order interactions and spontaneously broken supersymmetry}},
  \href{http://dx.doi.org/10.1103/PhysRevD.95.026003}{\emph{Phys. Rev.}
  {\bfseries D95} (2017) 026003},
  [\href{https://arxiv.org/abs/1601.03409}{{\ttfamily 1601.03409}}].

\bibitem{Vercnocke:2016fbt}
B.~Vercnocke and T.~Wrase, \emph{{Constrained superfields from an anti-D3-brane
  in KKLT}}, \href{http://dx.doi.org/10.1007/JHEP08(2016)132}{\emph{JHEP}
  {\bfseries 08} (2016) 132},
  [\href{https://arxiv.org/abs/1605.03961}{{\ttfamily 1605.03961}}].

\bibitem{Kallosh:2016aep}
R.~Kallosh, B.~Vercnocke and T.~Wrase, \emph{{String Theory Origin of
  Constrained Multiplets}},
  \href{http://dx.doi.org/10.1007/JHEP09(2016)063}{\emph{JHEP} {\bfseries 09}
  (2016) 063}, [\href{https://arxiv.org/abs/1606.09245}{{\ttfamily
  1606.09245}}].

\bibitem{Aalsma:2017ulu}
L.~Aalsma, J.~P. van~der Schaar and B.~Vercnocke, \emph{{Constrained
  superfields on metastable anti-D3-branes}},
  \href{http://dx.doi.org/10.1007/JHEP05(2017)089}{\emph{JHEP} {\bfseries 05}
  (2017) 089}, [\href{https://arxiv.org/abs/1703.05771}{{\ttfamily
  1703.05771}}].

\bibitem{Westphal:2006tn}
A.~Westphal, \emph{{de Sitter string vacua from Kahler uplifting}},
  \href{http://dx.doi.org/10.1088/1126-6708/2007/03/102}{\emph{JHEP} {\bfseries
  0703} (2007) 102}, [\href{https://arxiv.org/abs/hep-th/0611332}{{\ttfamily
  hep-th/0611332}}].

\bibitem{Sumitomo:2013vla}
Y.~Sumitomo, S.~H.~H. Tye and S.~S.~C. Wong, \emph{{Statistical Distribution of
  the Vacuum Energy Density in Racetrack K\"ahler Uplift Models in String
  Theory}}, \href{http://dx.doi.org/10.1007/JHEP07(2013)052}{\emph{JHEP}
  {\bfseries 07} (2013) 052},
  [\href{https://arxiv.org/abs/1305.0753}{{\ttfamily 1305.0753}}].

\bibitem{Achucarro:2006zf}
A.~Achucarro, B.~de~Carlos, J.~A. Casas and L.~Doplicher, \emph{{De Sitter
  vacua from uplifting D-terms in effective supergravities from realistic
  strings}}, \href{http://dx.doi.org/10.1088/1126-6708/2006/06/014}{\emph{JHEP}
  {\bfseries 06} (2006) 014},
  [\href{https://arxiv.org/abs/hep-th/0601190}{{\ttfamily hep-th/0601190}}].

\bibitem{Dudas:2006vc}
E.~Dudas and Y.~Mambrini, \emph{{Moduli stabilization with positive vacuum
  energy}}, \href{http://dx.doi.org/10.1088/1126-6708/2006/10/044}{\emph{JHEP}
  {\bfseries 10} (2006) 044},
  [\href{https://arxiv.org/abs/hep-th/0607077}{{\ttfamily hep-th/0607077}}].

\bibitem{Dudas:2006gr}
E.~Dudas, C.~Papineau and S.~Pokorski, \emph{{Moduli stabilization and
  uplifting with dynamically generated F-terms}},
  \href{http://dx.doi.org/10.1088/1126-6708/2007/02/028}{\emph{JHEP} {\bfseries
  02} (2007) 028}, [\href{https://arxiv.org/abs/hep-th/0610297}{{\ttfamily
  hep-th/0610297}}].

\bibitem{Abe:2006xp}
H.~Abe, T.~Higaki, T.~Kobayashi and Y.~Omura, \emph{{Moduli stabilization,
  F-term uplifting and soft supersymmetry breaking terms}},
  \href{http://dx.doi.org/10.1103/PhysRevD.75.025019}{\emph{Phys. Rev.}
  {\bfseries D75} (2007) 025019},
  [\href{https://arxiv.org/abs/hep-th/0611024}{{\ttfamily hep-th/0611024}}].

\bibitem{Parameswaran:2007kf}
S.~L. Parameswaran and A.~Westphal, \emph{{Consistent de Sitter string vacua
  from Kahler stabilization and D-term uplifting}},
  \href{http://dx.doi.org/10.1002/prop.200610374}{\emph{Fortsch. Phys.}
  {\bfseries 55} (2007) 804--810},
  [\href{https://arxiv.org/abs/hep-th/0701215}{{\ttfamily hep-th/0701215}}].

\bibitem{Lalak:2007qd}
Z.~Lalak, O.~J. Eyton-Williams and R.~Matyszkiewicz, \emph{{F-term uplifting
  via consistent D-terms}},
  \href{http://dx.doi.org/10.1088/1126-6708/2007/05/085}{\emph{JHEP} {\bfseries
  05} (2007) 085}, [\href{https://arxiv.org/abs/hep-th/0702026}{{\ttfamily
  hep-th/0702026}}].

\bibitem{Abe:2007yb}
H.~Abe, T.~Higaki and T.~Kobayashi, \emph{{More about F-term uplifting}},
  \href{http://dx.doi.org/10.1103/PhysRevD.76.105003}{\emph{Phys. Rev.}
  {\bfseries D76} (2007) 105003},
  [\href{https://arxiv.org/abs/0707.2671}{{\ttfamily 0707.2671}}].

\bibitem{Brax:2007xq}
P.~Brax, A.-C. Davis, S.~C. Davis, R.~Jeannerot and M.~Postma, \emph{{Warping
  and F-term uplifting}},
  \href{http://dx.doi.org/10.1088/1126-6708/2007/09/125}{\emph{JHEP} {\bfseries
  09} (2007) 125}, [\href{https://arxiv.org/abs/0707.4583}{{\ttfamily
  0707.4583}}].

\bibitem{Brax:2007fe}
P.~Brax, A.-C. Davis, S.~C. Davis, R.~Jeannerot and M.~Postma, \emph{{D-term
  Uplifted Racetrack Inflation}},
  \href{http://dx.doi.org/10.1088/1475-7516/2008/01/008}{\emph{JCAP} {\bfseries
  0801} (2008) 008}, [\href{https://arxiv.org/abs/0710.4876}{{\ttfamily
  0710.4876}}].

\bibitem{Dudas:2007nz}
E.~Dudas, Y.~Mambrini, S.~Pokorski and A.~Romagnoni, \emph{{Moduli
  stabilization with Fayet-Iliopoulos uplift}},
  \href{http://dx.doi.org/10.1088/1126-6708/2008/04/015}{\emph{JHEP} {\bfseries
  04} (2008) 015}, [\href{https://arxiv.org/abs/0711.4934}{{\ttfamily
  0711.4934}}].

\bibitem{Gallego:2008sv}
D.~Gallego and M.~Serone, \emph{{Moduli Stabilization in non-Supersymmetric
  Minkowski Vacua with Anomalous U(1) Symmetry}},
  \href{http://dx.doi.org/10.1088/1126-6708/2008/08/025}{\emph{JHEP} {\bfseries
  08} (2008) 025}, [\href{https://arxiv.org/abs/0807.0190}{{\ttfamily
  0807.0190}}].

\bibitem{Covi:2008zu}
L.~Covi, M.~Gomez-Reino, C.~Gross, G.~A. Palma and C.~A. Scrucca,
  \emph{{Constructing de Sitter vacua in no-scale string models without
  uplifting}},
  \href{http://dx.doi.org/10.1088/1126-6708/2009/03/146}{\emph{JHEP} {\bfseries
  0903} (2009) 146}, [\href{https://arxiv.org/abs/0812.3864}{{\ttfamily
  0812.3864}}].

\bibitem{Giveon:2009yu}
A.~Giveon, A.~Katz and Z.~Komargodski, \emph{{Uplifted Metastable Vacua and
  Gauge Mediation in SQCD}},
  \href{http://dx.doi.org/10.1088/1126-6708/2009/07/099}{\emph{JHEP} {\bfseries
  07} (2009) 099}, [\href{https://arxiv.org/abs/0905.3387}{{\ttfamily
  0905.3387}}].

\bibitem{Cicoli:2012fh}
M.~Cicoli, A.~Maharana, F.~Quevedo and C.~Burgess, \emph{{De Sitter String
  Vacua from Dilaton-dependent Non-perturbative Effects}},
  \href{http://dx.doi.org/10.1007/JHEP06(2012)011}{\emph{JHEP} {\bfseries 1206}
  (2012) 011}, [\href{https://arxiv.org/abs/1203.1750}{{\ttfamily 1203.1750}}].

\bibitem{Cicoli:2013rwa}
M.~Cicoli, S.~de~Alwis and A.~Westphal, \emph{{Heterotic Moduli
  Stabilisation}}, \href{http://dx.doi.org/10.1007/JHEP10(2013)199}{\emph{JHEP}
  {\bfseries 10} (2013) 199},
  [\href{https://arxiv.org/abs/1304.1809}{{\ttfamily 1304.1809}}].

\bibitem{Rummel:2014raa}
M.~Rummel and Y.~Sumitomo, \emph{{De Sitter Vacua from a D-term Generated
  Racetrack Uplift}},
  \href{http://dx.doi.org/10.1007/JHEP01(2015)015}{\emph{JHEP} {\bfseries 01}
  (2015) 015}, [\href{https://arxiv.org/abs/1407.7580}{{\ttfamily 1407.7580}}].

\bibitem{Kallosh:2014via}
R.~Kallosh and A.~Linde, \emph{{Inflation and Uplifting with Nilpotent
  Superfields}},
  \href{http://dx.doi.org/10.1088/1475-7516/2015/01/025}{\emph{JCAP} {\bfseries
  1501} (2015) 025}, [\href{https://arxiv.org/abs/1408.5950}{{\ttfamily
  1408.5950}}].

\bibitem{Cicoli:2015ylx}
M.~Cicoli, F.~Quevedo and R.~Valandro, \emph{{De Sitter from T-branes}},
  \href{http://dx.doi.org/10.1007/JHEP03(2016)141}{\emph{JHEP} {\bfseries 03}
  (2016) 141}, [\href{https://arxiv.org/abs/1512.04558}{{\ttfamily
  1512.04558}}].

\bibitem{Marsh:2014nla}
M.~C.~D. Marsh, B.~Vercnocke and T.~Wrase, \emph{{Decoupling and de Sitter
  Vacua in Approximate No-Scale Supergravities}},
  \href{http://dx.doi.org/10.1007/JHEP05(2015)081}{\emph{JHEP} {\bfseries 05}
  (2015) 081}, [\href{https://arxiv.org/abs/1411.6625}{{\ttfamily 1411.6625}}].

\bibitem{Kallosh:2014oja}
R.~Kallosh, A.~Linde, B.~Vercnocke and T.~Wrase, \emph{{Analytic Classes of
  Metastable de Sitter Vacua}},
  \href{http://dx.doi.org/10.1007/JHEP10(2014)011}{\emph{JHEP} {\bfseries 1410}
  (2014) 11}, [\href{https://arxiv.org/abs/1406.4866}{{\ttfamily 1406.4866}}].

\bibitem{Saltman:2004sn}
A.~Saltman and E.~Silverstein, \emph{{The Scaling of the no scale potential and
  de Sitter model building}},
  \href{http://dx.doi.org/10.1088/1126-6708/2004/11/066}{\emph{JHEP} {\bfseries
  0411} (2004) 066}, [\href{https://arxiv.org/abs/hep-th/0402135}{{\ttfamily
  hep-th/0402135}}].

\bibitem{Blaback:2015zra}
J.~{Bl\aa{}b\"ack}, U.~H. Danielsson, G.~Dibitetto and S.~C. Vargas,
  \emph{{Universal dS vacua in STU-models}},
  \href{http://dx.doi.org/10.1007/JHEP10(2015)069}{\emph{JHEP} {\bfseries 10}
  (2015) 069}, [\href{https://arxiv.org/abs/1505.04283}{{\ttfamily
  1505.04283}}].

\bibitem{Achucarro:2015kja}
A.~Achucarro, P.~Ortiz and K.~Sousa, \emph{{A new class of de Sitter vacua in
  String Theory Compactifications}},
  \href{http://dx.doi.org/10.1103/PhysRevD.94.086012}{\emph{Phys. Rev.}
  {\bfseries D94} (2016) 086012},
  [\href{https://arxiv.org/abs/1510.01273}{{\ttfamily 1510.01273}}].

\bibitem{Balasubramanian:2005zx}
V.~Balasubramanian, P.~Berglund, J.~P. Conlon and F.~Quevedo,
  \emph{{Systematics of moduli stabilisation in Calabi-Yau flux
  compactifications}},
  \href{http://dx.doi.org/10.1088/1126-6708/2005/03/007}{\emph{JHEP} {\bfseries
  0503} (2005) 007}, [\href{https://arxiv.org/abs/hep-th/0502058}{{\ttfamily
  hep-th/0502058}}].

\bibitem{Conlon:2005ki}
J.~P. Conlon, F.~Quevedo and K.~Suruliz, \emph{{Large-volume flux
  compactifications: Moduli spectrum and D3/D7 soft supersymmetry breaking}},
  \href{http://dx.doi.org/10.1088/1126-6708/2005/08/007}{\emph{JHEP} {\bfseries
  0508} (2005) 007}, [\href{https://arxiv.org/abs/hep-th/0505076}{{\ttfamily
  hep-th/0505076}}].

\bibitem{Ooguri:2016pdq}
H.~Ooguri and C.~Vafa, \emph{{Non-supersymmetric AdS and the Swampland}},
  \href{https://arxiv.org/abs/1610.01533}{{\ttfamily 1610.01533}}.

\bibitem{Freivogel:2016qwc}
B.~Freivogel and M.~Kleban, \emph{{Vacua Morghulis}},
  \href{https://arxiv.org/abs/1610.04564}{{\ttfamily 1610.04564}}.

\bibitem{Danielsson:2016mtx}
U.~Danielsson and G.~Dibitetto, \emph{{The fate of stringy AdS vacua and the
  WGC}},  \href{https://arxiv.org/abs/1611.01395}{{\ttfamily 1611.01395}}.

\bibitem{Grimm:2004uq}
T.~W. Grimm and J.~Louis, \emph{{The Effective action of N = 1 Calabi-Yau
  orientifolds}},
  \href{http://dx.doi.org/10.1016/j.nuclphysb.2004.08.005}{\emph{Nucl. Phys.}
  {\bfseries B699} (2004) 387--426},
  [\href{https://arxiv.org/abs/hep-th/0403067}{{\ttfamily hep-th/0403067}}].

\bibitem{Becker:2002nn}
K.~Becker, M.~Becker, M.~Haack and J.~Louis, \emph{{Supersymmetry breaking and
  alpha-prime corrections to flux induced potentials}},
  \href{http://dx.doi.org/10.1088/1126-6708/2002/06/060}{\emph{JHEP} {\bfseries
  06} (2002) 060}, [\href{https://arxiv.org/abs/hep-th/0204254}{{\ttfamily
  hep-th/0204254}}].

\bibitem{Berg:2007wt}
M.~Berg, M.~Haack and E.~Pajer, \emph{{Jumping Through Loops: On Soft Terms
  from Large Volume Compactifications}},
  \href{http://dx.doi.org/10.1088/1126-6708/2007/09/031}{\emph{JHEP} {\bfseries
  09} (2007) 031}, [\href{https://arxiv.org/abs/0704.0737}{{\ttfamily
  0704.0737}}].

\bibitem{Cicoli:2007xp}
M.~Cicoli, J.~P. Conlon and F.~Quevedo, \emph{{Systematics of String Loop
  Corrections in Type IIB Calabi-Yau Flux Compactifications}},
  \href{http://dx.doi.org/10.1088/1126-6708/2008/01/052}{\emph{JHEP} {\bfseries
  01} (2008) 052}, [\href{https://arxiv.org/abs/0708.1873}{{\ttfamily
  0708.1873}}].

\bibitem{Blumenhagen:2009gk}
R.~Blumenhagen, J.~P. Conlon, S.~Krippendorf, S.~Moster and F.~Quevedo,
  \emph{{SUSY Breaking in Local String/F-Theory Models}},
  \href{http://dx.doi.org/10.1088/1126-6708/2009/09/007}{\emph{JHEP} {\bfseries
  09} (2009) 007}, [\href{https://arxiv.org/abs/0906.3297}{{\ttfamily
  0906.3297}}].

\bibitem{Gukov:1999ya}
S.~Gukov, C.~Vafa and E.~Witten, \emph{{CFT's from Calabi-Yau four folds}},
  \href{http://dx.doi.org/10.1016/S0550-3213(01)00289-9,
  10.1016/S0550-3213(00)00373-4}{\emph{Nucl. Phys.} {\bfseries B584} (2000)
  69--108}, [\href{https://arxiv.org/abs/hep-th/9906070}{{\ttfamily
  hep-th/9906070}}].

\bibitem{Grana:2000jj}
M.~Grana and J.~Polchinski, \emph{{Supersymmetric three form flux perturbations
  on AdS(5)}}, \href{http://dx.doi.org/10.1103/PhysRevD.63.026001}{\emph{Phys.
  Rev.} {\bfseries D63} (2001) 026001},
  [\href{https://arxiv.org/abs/hep-th/0009211}{{\ttfamily hep-th/0009211}}].

\bibitem{Gubser:2000vg}
S.~S. Gubser, \emph{{Supersymmetry and F theory realization of the deformed
  conifold with three form flux}},
  \href{https://arxiv.org/abs/hep-th/0010010}{{\ttfamily hep-th/0010010}}.

\bibitem{Grana:2001xn}
M.~Grana and J.~Polchinski, \emph{{Gauge / gravity duals with holomorphic
  dilaton}}, \href{http://dx.doi.org/10.1103/PhysRevD.65.126005}{\emph{Phys.
  Rev.} {\bfseries D65} (2002) 126005},
  [\href{https://arxiv.org/abs/hep-th/0106014}{{\ttfamily hep-th/0106014}}].

\bibitem{Candelas:1990pi}
P.~Candelas and X.~de~la Ossa, \emph{{Moduli Space of {Calabi-Yau} Manifolds}},
  \href{http://dx.doi.org/10.1016/0550-3213(91)90122-E}{\emph{Nucl. Phys.}
  {\bfseries B355} (1991) 455--481}.

\bibitem{Brodie:2015kza}
C.~Brodie and M.~C.~D. Marsh, \emph{{The Spectra of Type IIB Flux
  Compactifications at Large Complex Structure}},
  \href{http://dx.doi.org/10.1007/JHEP01(2016)037}{\emph{JHEP} {\bfseries 01}
  (2016) 037}, [\href{https://arxiv.org/abs/1509.06761}{{\ttfamily
  1509.06761}}].

\bibitem{Marsh:2015zoa}
M.~C.~D. Marsh and K.~Sousa, \emph{{Universal Properties of Type IIB and
  F-theory Flux Compactifications at Large Complex Structure}},
  \href{http://dx.doi.org/10.1007/JHEP03(2016)064}{\emph{JHEP} {\bfseries 03}
  (2016) 064}, [\href{https://arxiv.org/abs/1512.08549}{{\ttfamily
  1512.08549}}].

\bibitem{Cremmer:1983bf}
E.~Cremmer, S.~Ferrara, C.~Kounnas and D.~V. Nanopoulos, \emph{{Naturally
  Vanishing Cosmological Constant in N=1 Supergravity}},
  \href{http://dx.doi.org/10.1016/0370-2693(83)90106-5}{\emph{Phys.Lett.}
  {\bfseries B133} (1983) 61}.

\bibitem{Ellis:1983ei}
J.~R. Ellis, C.~Kounnas and D.~V. Nanopoulos, \emph{{Phenomenological SU(1,1)
  Supergravity}},
  \href{http://dx.doi.org/10.1016/0550-3213(84)90054-3}{\emph{Nucl. Phys.}
  {\bfseries B241} (1984) 406--428}.

\bibitem{Ellis:1984bm}
J.~R. Ellis, C.~Kounnas and D.~V. Nanopoulos, \emph{{No Scale Supersymmetric
  Guts}}, \href{http://dx.doi.org/10.1016/0550-3213(84)90555-8}{\emph{Nucl.
  Phys.} {\bfseries B247} (1984) 373--395}.

\bibitem{Danielsson:2015rca}
U.~Danielsson and G.~Dibitetto, \emph{{Type IIB on $S^{3}\times S^{3}$ through
  $Q$ \& $P$ fluxes}},
  \href{http://dx.doi.org/10.1007/JHEP01(2016)057}{\emph{JHEP} {\bfseries 01}
  (2016) 057}, [\href{https://arxiv.org/abs/1507.04476}{{\ttfamily
  1507.04476}}].

\bibitem{Gallego:2011jm}
D.~Gallego, \emph{{On the Effective Description of Large Volume
  Compactifications}},
  \href{http://dx.doi.org/10.1007/JHEP06(2011)087}{\emph{JHEP} {\bfseries 06}
  (2011) 087}, [\href{https://arxiv.org/abs/1103.5469}{{\ttfamily 1103.5469}}].

\bibitem{Gallego:2013qe}
D.~Gallego, \emph{{Light field integration in SUGRA theories}},
  \href{http://dx.doi.org/10.1142/S0217751X15500037}{\emph{Int. J. Mod. Phys.}
  {\bfseries A30} (2015) 1550003},
  [\href{https://arxiv.org/abs/1301.6177}{{\ttfamily 1301.6177}}].

\bibitem{Sousa:2014qza}
K.~Sousa and P.~Ortiz, \emph{{Perturbative Stability along the Supersymmetric
  Directions of the Landscape}},
  \href{http://dx.doi.org/10.1088/1475-7516/2015/02/017}{\emph{JCAP} {\bfseries
  1502} (2015) 017}, [\href{https://arxiv.org/abs/1408.6521}{{\ttfamily
  1408.6521}}].

\bibitem{Dine:1985he}
M.~Dine and N.~Seiberg, \emph{{Is the Superstring Weakly Coupled?}},
  \href{http://dx.doi.org/10.1016/0370-2693(85)90927-X}{\emph{Phys. Lett.}
  {\bfseries B162} (1985) 299--302}.

\bibitem{Denef:2004dm}
F.~Denef, M.~R. Douglas and B.~Florea, \emph{{Building a better racetrack}},
  \href{http://dx.doi.org/10.1088/1126-6708/2004/06/034}{\emph{JHEP} {\bfseries
  06} (2004) 034}, [\href{https://arxiv.org/abs/hep-th/0404257}{{\ttfamily
  hep-th/0404257}}].

\bibitem{Kachru:2003aw}
S.~Kachru, R.~Kallosh, A.~D. Linde and S.~P. Trivedi, \emph{{De Sitter vacua in
  string theory}},
  \href{http://dx.doi.org/10.1103/PhysRevD.68.046005}{\emph{Phys.Rev.}
  {\bfseries D68} (2003) 046005},
  [\href{https://arxiv.org/abs/hep-th/0301240}{{\ttfamily hep-th/0301240}}].

\bibitem{Krippendorf:2010hj}
S.~Krippendorf, M.~J. Dolan, A.~Maharana and F.~Quevedo, \emph{{D-branes at
  Toric Singularities: Model Building, Yukawa Couplings and Flavour Physics}},
  \href{http://dx.doi.org/10.1007/JHEP06(2010)092}{\emph{JHEP} {\bfseries 06}
  (2010) 092}, [\href{https://arxiv.org/abs/1002.1790}{{\ttfamily 1002.1790}}].

\bibitem{Dolan:2011qu}
M.~J. Dolan, S.~Krippendorf and F.~Quevedo, \emph{{Towards a Systematic
  Construction of Realistic D-brane Models on a del Pezzo Singularity}},
  \href{http://dx.doi.org/10.1007/JHEP10(2011)024}{\emph{JHEP} {\bfseries 10}
  (2011) 024}, [\href{https://arxiv.org/abs/1106.6039}{{\ttfamily 1106.6039}}].

\bibitem{Cicoli:2012vw}
M.~Cicoli, S.~Krippendorf, C.~Mayrhofer, F.~Quevedo and R.~Valandro,
  \emph{{D-Branes at del Pezzo Singularities: Global Embedding and Moduli
  Stabilisation}}, \href{http://dx.doi.org/10.1007/JHEP09(2012)019}{\emph{JHEP}
  {\bfseries 1209} (2012) 019},
  [\href{https://arxiv.org/abs/1206.5237}{{\ttfamily 1206.5237}}].

\bibitem{Cicoli:2013cha}
M.~Cicoli, D.~Klevers, S.~Krippendorf, C.~Mayrhofer, F.~Quevedo et~al.,
  \emph{{Explicit de Sitter Flux Vacua for Global String Models with Chiral
  Matter}}, \href{http://dx.doi.org/10.1007/JHEP05(2014)001}{\emph{JHEP}
  {\bfseries 1405} (2014) 001},
  [\href{https://arxiv.org/abs/1312.0014}{{\ttfamily 1312.0014}}].

\bibitem{Cicoli:2013mpa}
M.~Cicoli, S.~Krippendorf, C.~Mayrhofer, F.~Quevedo and R.~Valandro,
  \emph{{D3/D7 Branes at Singularities: Constraints from Global Embedding and
  Moduli Stabilisation}},
  \href{http://dx.doi.org/10.1007/JHEP07(2013)150}{\emph{JHEP} {\bfseries 07}
  (2013) 150}, [\href{https://arxiv.org/abs/1304.0022}{{\ttfamily 1304.0022}}].

\bibitem{Blumenhagen:2007sm}
R.~Blumenhagen, S.~Moster and E.~Plauschinn, \emph{{Moduli Stabilisation versus
  Chirality for MSSM like Type IIB Orientifolds}},
  \href{http://dx.doi.org/10.1088/1126-6708/2008/01/058}{\emph{JHEP} {\bfseries
  01} (2008) 058}, [\href{https://arxiv.org/abs/0711.3389}{{\ttfamily
  0711.3389}}].

\bibitem{Kaplunovsky:1993rd}
V.~S. Kaplunovsky and J.~Louis, \emph{{Model independent analysis of soft terms
  in effective supergravity and in string theory}},
  \href{http://dx.doi.org/10.1016/0370-2693(93)90078-V}{\emph{Phys. Lett.}
  {\bfseries B306} (1993) 269--275},
  [\href{https://arxiv.org/abs/hep-th/9303040}{{\ttfamily hep-th/9303040}}].

\bibitem{Brignole:1993dj}
A.~Brignole, L.~E. Ibanez and C.~Munoz, \emph{{Towards a theory of soft terms
  for the supersymmetric Standard Model}},
  \href{http://dx.doi.org/10.1016/0550-3213(94)00600-J,
  10.1016/0550-3213(94)00068-9}{\emph{Nucl. Phys.} {\bfseries B422} (1994)
  125--171}, [\href{https://arxiv.org/abs/hep-ph/9308271}{{\ttfamily
  hep-ph/9308271}}].

\bibitem{Soni:1983rm}
S.~K. Soni and H.~A. Weldon, \emph{{Analysis of the Supersymmetry Breaking
  Induced by N=1 Supergravity Theories}},
  \href{http://dx.doi.org/10.1016/0370-2693(83)90593-2}{\emph{Phys. Lett.}
  {\bfseries 126B} (1983) 215--219}.

\bibitem{Conlon:2006tj}
J.~P. Conlon, D.~Cremades and F.~Quevedo, \emph{{Kahler potentials of chiral
  matter fields for Calabi-Yau string compactifications}},
  \href{http://dx.doi.org/10.1088/1126-6708/2007/01/022}{\emph{JHEP} {\bfseries
  01} (2007) 022}, [\href{https://arxiv.org/abs/hep-th/0609180}{{\ttfamily
  hep-th/0609180}}].

\bibitem{Berg:2010ha}
M.~Berg, D.~Marsh, L.~McAllister and E.~Pajer, \emph{{Sequestering in String
  Compactifications}},
  \href{http://dx.doi.org/10.1007/JHEP06(2011)134}{\emph{JHEP} {\bfseries 06}
  (2011) 134}, [\href{https://arxiv.org/abs/1012.1858}{{\ttfamily 1012.1858}}].

\bibitem{Randall:1998uk}
L.~Randall and R.~Sundrum, \emph{{Out of this world supersymmetry breaking}},
  \href{http://dx.doi.org/10.1016/S0550-3213(99)00359-4}{\emph{Nucl. Phys.}
  {\bfseries B557} (1999) 79--118},
  [\href{https://arxiv.org/abs/hep-th/9810155}{{\ttfamily hep-th/9810155}}].

\bibitem{Aparicio:2014wxa}
L.~Aparicio, M.~Cicoli, S.~Krippendorf, A.~Maharana, F.~Muia and F.~Quevedo,
  \emph{{Sequestered de Sitter String Scenarios: Soft-terms}},
  \href{http://dx.doi.org/10.1007/JHEP11(2014)071}{\emph{JHEP} {\bfseries 11}
  (2014) 071}, [\href{https://arxiv.org/abs/1409.1931}{{\ttfamily 1409.1931}}].

\bibitem{Berg:2012aq}
M.~Berg, J.~P. Conlon, D.~Marsh and L.~T. Witkowski, \emph{{Superpotential
  de-sequestering in string models}},
  \href{http://dx.doi.org/10.1007/JHEP02(2013)018}{\emph{JHEP} {\bfseries 02}
  (2013) 018}, [\href{https://arxiv.org/abs/1207.1103}{{\ttfamily 1207.1103}}].

\bibitem{Ibanez:1998rf}
L.~E. Ibanez, C.~Munoz and S.~Rigolin, \emph{{Aspect of type I string
  phenomenology}},
  \href{http://dx.doi.org/10.1016/S0550-3213(99)00264-3}{\emph{Nucl. Phys.}
  {\bfseries B553} (1999) 43--80},
  [\href{https://arxiv.org/abs/hep-ph/9812397}{{\ttfamily hep-ph/9812397}}].

\bibitem{Lust:2004cx}
D.~Lust, P.~Mayr, R.~Richter and S.~Stieberger, \emph{{Scattering of gauge,
  matter, and moduli fields from intersecting branes}},
  \href{http://dx.doi.org/10.1016/j.nuclphysb.2004.06.052}{\emph{Nucl. Phys.}
  {\bfseries B696} (2004) 205--250},
  [\href{https://arxiv.org/abs/hep-th/0404134}{{\ttfamily hep-th/0404134}}].

\bibitem{Lust:2004fi}
D.~Lust, S.~Reffert and S.~Stieberger, \emph{{Flux-induced soft supersymmetry
  breaking in chiral type IIB orientifolds with D3 / D7-branes}},
  \href{http://dx.doi.org/10.1016/j.nuclphysb.2004.11.030}{\emph{Nucl. Phys.}
  {\bfseries B706} (2005) 3--52},
  [\href{https://arxiv.org/abs/hep-th/0406092}{{\ttfamily hep-th/0406092}}].

\bibitem{Coughlan:1983ci}
G.~D. Coughlan, W.~Fischler, E.~W. Kolb, S.~Raby and G.~G. Ross,
  \emph{{Cosmological Problems for the Polonyi Potential}},
  \href{http://dx.doi.org/10.1016/0370-2693(83)91091-2}{\emph{Phys. Lett.}
  {\bfseries B131} (1983) 59--64}.

\bibitem{Banks:1993en}
T.~Banks, D.~B. Kaplan and A.~E. Nelson, \emph{{Cosmological implications of
  dynamical supersymmetry breaking}},
  \href{http://dx.doi.org/10.1103/PhysRevD.49.779}{\emph{Phys. Rev.} {\bfseries
  D49} (1994) 779--787},
  [\href{https://arxiv.org/abs/hep-ph/9308292}{{\ttfamily hep-ph/9308292}}].

\bibitem{deCarlos:1993wie}
B.~de~Carlos, J.~A. Casas, F.~Quevedo and E.~Roulet, \emph{{Model independent
  properties and cosmological implications of the dilaton and moduli sectors of
  4-d strings}},
  \href{http://dx.doi.org/10.1016/0370-2693(93)91538-X}{\emph{Phys. Lett.}
  {\bfseries B318} (1993) 447--456},
  [\href{https://arxiv.org/abs/hep-ph/9308325}{{\ttfamily hep-ph/9308325}}].

\end{thebibliography}\endgroup
	
\end{document}